\documentclass{aa}  
\bibpunct{(}{)}{;}{a}{}{,} 
\usepackage{hyperref}
\usepackage{graphicx}
\usepackage{txfonts}
\usepackage{natbib}

\newcommand{\be}{\begin{equation}}
\newcommand{\ee}{\end{equation}}
\newcommand{\bea}{\begin{eqnarray}}
\newcommand{\eea}{\end{eqnarray}}

\usepackage{ulem}

\begin{document}

\title{Inference of neutron-star properties with unified crust-core equations of state for parameter estimation}

\author{P. J. Davis\inst{1}, H. Dinh Thi\inst{1}, A. F. Fantina\inst{2}, F. Gulminelli\inst{1}, M. Oertel\inst{3}, L. Suleiman\inst{3,4,5}}

\institute{Université de Caen Normandie, ENSICAEN, CNRS/IN2P3, LPC Caen UMR6534, F-14000 Caen, France \\
\email{davis@lpccaen.in2p3.fr} 
\and Grand Acc\'el\'erateur National d'Ions Lourds (GANIL), CEA/DRF - CNRS/IN2P3, Boulevard Henri Becquerel, 14076 Caen, France 
\and Laboratoire Univers et Th\'eories, CNRS, Observatoire de Paris, Universit\'e PSL, Université Paris Cité, 5 place Jules Janssen, 92195 Meudon, France 
\and Nicolaus Copernicus Astronomical Center, Polish Academy of Sciences, Bartycka 18, PL-00-716 Warszawa, Poland 
\and Nicholas and Lee Begovich Center for Gravitational Wave Physics and Astronomy, California State University Fullerton, Fullerton, California 92831, USA.
}


\titlerunning{Unified equations of state for parameter estimations}
\authorrunning{Davis et al.}

\abstract
   {Relating different global neutron-star (NS) properties, such as tidal deformability and radius, or mass and radius, requires an equation of state (EoS). Determining the NS EoS is therefore not only the science goal of a variety of observational projects, but it also enters in the analysis process; for example, to predict a NS radius from a measured tidal deformability via gravitational waves (GW) during the inspiral of a binary NS merger. To this aim, it is important to estimate the theoretical uncertainties on the EoS, one of which is the possible bias coming from an inconsistent treatment of the low-density region; that is, the use of a so called non-unified NS crust. 
   }
     {We propose a numerical tool allowing the user to consistently match a nuclear-physics informed crust to an arbitrary high-density EoS describing the core of the star.}
   {We introduce an inversion procedure of the EoS close to saturation density that allows users to extract nuclear-matter parameters and extend the EoS to lower densities in a consistent way. For the treatment of inhomogeneous matter in the crust, a standard approach based on the compressible liquid-drop (CLD) model approach was used in our work. 
   A Bayesian analysis using a parametric agnostic EoS representation in the high-density region is also presented in order to quantify the uncertainties induced by an inconsistent treatment of the crust.} 
   {We show that the use of a fixed, realistic-but-inconsistent model for the crust causes small but avoidable errors in the estimation of global NS properties and leads to an underestimation of the uncertainties in the inference of NS properties. }
   {Our results highlight the importance of employing a consistent EoS in inference schemes. The numerical tool that we developed to reconstruct such a thermodynamically consistent EoS, CUTER, has been tested and validated for use by the astrophysical community. } 


\maketitle


\section{Introduction}
\label{sec:intro}

The detection of the first binary neutron-star (NS) merger by the LIGO-Virgo collaboration (LVC), the event GW170817 \citep{Abbott2017a}, and the associated electromagnetic counterpart \citep{Abbott2017b, Abbott2017c} has opened a new window to explore matter under extreme conditions.
In particular, the imprint on the gravitational-wave (GW) signal left by the tidal interaction of the two NSs during the pre-merger phase led to the first estimate of the tidal deformability \citep{Abbott2018prlb, Abbott2019prx}, which already made it possible to put some constraints on the properties of dense matter (see \citet{Chatziioannou2020} and references therein for a review).
Together with the precise mass determinations of three massive pulsars with masses around $2 M_\odot$ --- PSR J1614-2230~\citep{Demorest2010}, PSR J0348$+$0432~\citep{Antoniadis2013}, and PSR J0740+6620~\citep{Cromartie2020} --- as well as the recent quantitative estimations of the mass-radius relation of PSR J0740$+$6620 and PSR J0030$+$0451 obtained from the Neutron-star Interior Composition ExploreR (NICER) measurements \citep{Riley2019,Riley2021, Raaijmakers2019, Miller2019, Miller2021}, these observations advanced the study of the properties of dense matter in NSs.
Moreover, the plethora of upcoming observations from the LIGO-Virgo-KAGRA collaboration, as well as those foreseen from the planned third-generation detectors such as the Einstein Telescope \citep{Maggiore2020, Branchesi2023} or the Cosmic Explorer \citep{Evans2021}, are expected to provide valuable data from which new information on the structure and composition of NSs can be drawn.

In order to reliably relate different astrophysical observations and extract information both on NS properties and on the dense-matter properties in the interior of NSs, a proper description of the NS equation of state (EoS), that is, the functional relation between the pressure and the mass-energy density, is needed. 
By `proper description' we mean a consistent mapping of the underlying interaction model on all parts of the internal structure of the star.
Indeed, under the assumption of general relativity, the connection between static properties of cold (beta-equilibrated) NSs (which is a valid assumption for mature isolated NSs or coalescing NSs in their inspiral phase such as those considered in this manuscript) and the underlying microphysics mainly relies on the knowledge of the EoS.
Most current inferences of NS properties from astrophysical (GW) data are done using agnostic EoSs, that is, EoSs that are not informed by a specific description of the microphysics (they can be parametric --- such as piecewise polytropes or sound-speed models --- or non-parametric --- such as Gaussian processes).
These EoSs are usually matched with a (unique) given crust at low densities (typically, the crust of \citet{Baym1971} or of \citet{Douchin01} is used; see e.g. \citet{Essick2020, Essick2021, Greif2019, Huang2023, Landry2019, Raithel2023}).
The reason for this choice is based on the idea that the (low-density) crust EoS is relatively well constrained and that the latter has a negligible effect on NS macroscopic observables, which are instead mainly determined by the largely unknown high-density part of the EoS. 
However, the determination of the crust EoS, particularly in the inner crust, still suffers from model dependencies (see e.g. \citet{Haensel2007, Oertel2017, Burgio2018} for a review).
In addition, recent studies have shown that an incorrect treatment of the crust and the way the crust is matched to the core EoS can lead to spurious errors in the prediction of the NS macroscopic properties from a given EoS model (see e.g. \cite{Fortin2016, Ferreira2020, Suleiman2021, Suleiman2022}). 
These uncertainties are small compared with current precision of astronomical data, but they will become relevant in the near future, in particular for third-generation GW detectors where, for example, the NS radius is expected to be determined to less than $1\%$ precision (see e.g.~\citet{Chatziioannou2021, Huxford2023,Iacovelli2023}). 
Therefore, it is timely to provide thermodynamically consistent and unified (meaning that the same nuclear model is employed for the different regions of the NSs) EoSs for inference analyses.
This is indeed a difficult task since the interior of a NS comprises many different structures from the outer crust with a crystalline structure of ions immersed in an electron gas up to the homogeneous dense core, where large uncertainties on the composition still remain. 
Each structure thereby needs a specific treatment requiring a consistent matching to obtain a unified description of the entire NS interior.

The large uncertainties at high density in the NS core arise since degrees of freedom other than nucleonic ones can emerge, and phase transitions to quark matter may even occur (see e.g. \citet{Haensel2007, Blaschke2018, Oertel2017, Raduta2022} for a review); thus, agnostic models are often used in inference schemes to account for our incomplete knowledge of the high-density EoS.
Such phase transitions, however, are unlikely to take place before nuclear saturation density\footnote{The nuclear saturation density, $n_{\rm sat} \sim 0.16$~$\mathrm{fm}^{-3}$, is defined as the density at which the energy per baryon of symmetric (i.e. equal number of protons and neutrons) nuclear matter has a minimum.} (and even twice saturation density), where only nucleonic degrees of freedom are expected to be present (see e.g. \citet{Tang2021, Christian2020}).
Therefore, a possible way to build a consistent EoS without losing any freedom in the description of the high-density EoS is to extract the nuclear parameters from the high-density EoS around saturation density, similarly to the procedure proposed in \citet{Essick2021}, and subsequently calculate the low-density EoS with the very same parameters. 
Based on this idea, in this work we propose an efficient way to construct consistent and unified crust--core EoSs of cold and beta-equilibrated NSs for astrophysical (in particular GW signal) analysis. 

The formalism is presented in Sect.~\ref{sec:method}, in which we first discuss in Sect.~\ref{sec:inversion} how to extract the nuclear parameters and reconstruct a consistent EoS starting from a relation between the baryon number density and the energy density in beta equilibrium, while we specifically present the framework to calculate the crust EoS in Sect.~\ref{sec:crust} and the global NS properties in Sect.~\ref{sec:NS-prop}.
Numerical results are given in Sect.~\ref{sec:results}, in which we first discuss in Sect.~\ref{sec:code-valid} the procedure and code validation and show the performance of the EoS reconstruction for a limited set of models.
To assess the uncertainties associated with the use of a unique crust EoS instead of a consistent model, we perform a Bayesian analysis and systematically compare the predictions of different NS observables predicted by two scenarios: (1) those obtained using a set of crust EoSs matched to (agnostic) polytropic EoSs at high density, and (2) those obtained by employing a unique EoS model at lower densities; results are shown in Sect.~\ref{sec:bayes}.
Finally, we draw our conclusions in Sect.~\ref{sec:conclusions}.

\section{Consistent equation-of-state reconstruction}
\label{sec:method}

\subsection{Extracting the nuclear-matter parameters from the neutron-star equation of state}
\label{sec:inversion}

Let us suppose that the relation between the energy density of cold beta-equilibrated NS matter $\mathcal{E}_\beta$ and the baryon number density $n_B$ in the zero-temperature approximation is given by an input EoS, not necessarily associated with a specific microscopic model.
At a baryonic density low enough that a nucleonic content can safely be assumed (typically less than the saturation density, which corresponds to the internal density of terrestrial atomic nuclei), it is possible to define the energy per baryon of homogeneous nucleonic (neutron, proton, and electron) matter that is consistent with the assumed $\mathcal{E}_\beta$ as\footnote{We use the uppercase $E$, the lowercase $e$, and $\mathcal{E}$ to denote the total energy, the energy per nucleon, and the energy per unit volume, respectively.}
\begin{eqnarray}
e_{\rm nuc,\beta}(n_B) &=& \frac{1}{n_B}\left [ \mathcal{E}_\beta(n_B) - \mathcal{E}_e(n_e(n_B)) \right. \nonumber \\
&& - \left. n_e m_p c^2 - (n_B - n_e) m_n c^2 \right] \ ,
\label{eq:enuc}
\end{eqnarray}
with $m_n$ and $m_p$ being the neutron and proton mass\footnote{We note that the expression for the nucleon masses that we adopted for the subtraction in Eq.~\eqref{eq:enuc} (see also Eq.~\eqref{eq:edens-meta}) might not be exactly the same as that employed in the original EoS, $\mathcal{E}_\beta$. This can in principle lead to a small inconsistency that cannot be avoided if details on the mass definition in the original EoS model are not known.}, respectively, $c$ the speed of light, and $n_e$ the electron density that corresponds to the condition of equilibrium of weak reactions in the nucleonic regime:
\begin{equation}
\mu_n(n_B)-\mu_p(n_B)=\mu_e(n_e) \ .
\label{eq:beta}
\end{equation}
The electron energy density is denoted $\mathcal{E}_e$ and given by the expression for an ideal Fermi gas,
\begin{equation}
\mathcal{E}_e=\frac{(m_e c^2)^4}{8 \pi^2 (\hbar c)^3}\left [ x_r(2x_r^2+1)\sqrt{x_r^2+1}- \ln \left( x_r + \sqrt{x_r^2+1} \right) \right ] \;,
\label{eq:e_el}
\end{equation}
with $x_r= \hbar c (3\pi^2 n_e)^{1/3}/(m_e c^2)$ being the relativity parameter, $\hbar$ the Dirac-Planck constant, and $m_e$ the electron mass (see e.g. \citet{Weiss2004}). 
We note that in principle muons can contribute to the EoS, thus to $\mathcal{E}_\beta$. However, in Eq.~\eqref{eq:enuc} we only consider electrons since muons are not expected to contribute before saturation density; therefore, the procedure of the extraction of the nuclear-matter parameters described here remains valid.
The isospin asymmetry parameter is denoted as $\delta$ and defined using the neutron and proton number density, $n_n$ and $n_p$, respectively, as $\delta = (n_n-n_p)/n_B$. 
For a pure nucleonic content, the value of the isospin asymmetry at each density, $\delta_{\beta}(n_B)$, is related to the electron number density by the charge neutrality condition $n_e=n_B\,(1-\delta_\beta)/2$.

In order to build a consistent description of the crust with standard variational tools, as pioneered by \citet{Baym1971} and \citet{Douchin01}, one needs to know the behaviour of $e_{\rm nuc}$ for different values of $0<\delta<1$ besides the $\beta$-equilibrium value $\delta_\beta(n_B) $ in the sub-saturation density regime.
To this aim, we write the energy density of nucleonic matter using the following expression, as in the meta-model approach developed in \citet{Margueron2018}: 
\begin{equation}
 \mathcal{E}_{\rm nuc}(n_B,\delta) = 
 \sum_{q=n,p}n_q \, m_qc^2  + n_B e_{\rm nuc}(n_B,\delta)  \ ,
 \label{eq:edens-meta}
\end{equation}
where $q=n,p$ labels neutron and proton, respectively, and the energy per baryon is given by
\begin{equation}
e_{\rm nuc}(n_B,\delta) = t_{\rm FG}^\star(n_B,\delta) +e_{\rm is}(n_B)+e_{\rm iv}(n_B)\delta^2 \ ,
\label{eq:meta}
\end{equation}
where $e_{\rm is}(n_B)$ ($e_{\rm iv}(n_B)$) is the isoscalar (isovector) energy (see Eq.~\eqref{eq:e0} and the following).
We note that Eq.~\eqref{eq:meta} is not a purely parabolic approximation for the isospin dependence because of the kinetic term $t_{\rm FG}^\star$, which includes the dominant deviation to the parabolic approximation,
\begin{eqnarray}
t_{\rm FG}^\star(n,\delta) &=& \frac{t_{\rm FG, sat}}{2}\left(\frac{n_B}{n_{\rm sat}}\right)^{2/3} 
\bigg[ \left( 1+\kappa_{\rm sat}\frac{n_B}{n_{\rm sat}} \right) f_1(\delta) \nonumber \\
&& + \kappa_{\rm sym}\frac{n_B}{n_{\rm sat}}f_2(\delta)\bigg] ,
\label{eq:effmassms2}
\end{eqnarray}
where $t_{\rm FG, sat} = 3 \hbar^2 c^2 \left(3\pi^{2}/2\right)^{2/3}n_{\rm sat}^{2/3}/(10 m c^2) $ is the kinetic energy per nucleon of symmetric nuclear matter at the saturation density $n_{\rm sat}$, with $m=(m_n+m_p)/2$, 
and the functions $f_i$ are defined as
\bea
f_1(\delta) &=& (1+\delta)^{5/3}+(1-\delta)^{5/3}  \;, \\
f_2(\delta) &=& \delta \left[ (1+\delta)^{5/3}-(1-\delta)^{5/3} \right] \ .
\eea
The terms $\kappa_{\rm sat}$ and $\kappa_{\rm sym}$ in Eq.~(\ref{eq:effmassms2}) are linked to the nucleon effective masses $m_q^\star$ by 
\begin{eqnarray}
\kappa_{\rm sat}&=&\frac{m}{m_{\rm sat}^\star} - 1 \;\;  \hbox{ in symmetric matter ($\delta=0$)}  \;, \nonumber \\
\kappa_{\rm sym}&=&\frac 1 2 \left[ \frac{m}{m^\star_n} - \frac{m}{m^\star_p}  \right] \hbox{ in neutron matter ($\delta=1$)} \ \ ,
\end{eqnarray}
where $m_{\rm sat}^\star = m^\star(n_{\rm sat})$ is the Landau effective mass at saturation (see also the discussion in \citet{Margueron2018}).
In order to equate Eqs.~\eqref{eq:enuc} and \eqref{eq:meta} along the beta-equilibrium curve $\delta=\delta_\beta(n_B)$, the chemical potentials have to be computed.
The electron chemical potential is simply given by its Fermi energy,
\begin{equation}
\mu_e = \sqrt{(\hbar c)^2(3\pi^2n_e)^{2/3} + (m_e c^2)^2} \;,
\end{equation}
and the neutron and proton chemical potentials are given by
\begin{equation}
\label{eq:muq}
\mu_q=\left. \frac{\partial \mathcal{E}_{\rm nuc}}{\partial n_q} \right|_{n_{q^\prime}} 
= e_{\rm nuc}+m_q c^2 + \left. n_B \frac{\partial e_{\rm nuc}}{\partial n_B} \right|_\delta +(-\delta \pm 1) \left. \frac{\partial e_{\rm nuc}}{\partial \delta}\right|_{n_B}, 
\end{equation}
where the plus (minus) sign holds for neutrons (protons).
Replacing Eq.~(\ref{eq:muq}) in Eq.~(\ref{eq:beta}), we obtain the standard beta-equilibrium condition (see \citet{Lattimer1991prl, Li1998}):
\begin{equation}
 2 \left. \frac{\partial e_{\rm nuc}}{\partial \delta}\right|_{n_B}(n_B,\delta_\beta) = \mu_e(n_B,\delta_\beta)-\Delta m_{np} c^2 \ ,
\end{equation}
with $\Delta m_{np}=m_n-m_p$, 
or equivalently, using the expansion in Eq.~\eqref{eq:meta},
\begin{equation}
2 \left. \frac{\partial t_{\rm FG}^\star}{\partial \delta}\right|_{n_B} (n_B,\delta_\beta)-\mu_e(n_B,\delta_\beta)+\Delta m_{np} c^2 
= - 4\delta_\beta e_{\rm iv}(n_B) \;. \label{eq:beta1}
\end{equation}
The unknown density-dependent isovector energy $e_{\rm iv}(n_B)$ in Eq.~(\ref{eq:beta1}) can be extracted from the input $\mathcal{E}_\beta(n_B)$ using Eqs.~(\ref{eq:enuc}) and (\ref{eq:meta}):
\begin{eqnarray}
\delta_\beta^2 e_{\rm iv}(n_B) & = &\frac{1}{n_B}\left [ \mathcal{E}_\beta(n_B) - \mathcal{E}_e(n_e) \right. \nonumber \\
&& - \left. n_e m_p c^2 - (n_B - n_e) m_n c^2 \right] \nonumber \\ && - t_{\rm FG}^\star(n_B,\delta_\beta) - e_{\rm is}(n_B). 
\label{esym}
\end{eqnarray}
Replacing Eq.~(\ref{esym}) in Eq.~(\ref{eq:beta1}), we have
\begin{eqnarray}
&& 2 \left. \frac{\partial t_{\rm FG}^\star}{\partial \delta}\right|_{n_B}(n_B,\delta_\beta) - \mu_e(n_B,\delta_\beta) + \Delta m_{np} c^2\label{final}  \\
&& = -\frac{4}{n_B \delta_\beta}\left [ \mathcal{E}_\beta(n_B) - \mathcal{E}_e(n_e) \right. \nonumber \\ 
&& - n_e m_p c^2 - (n_B - n_e) m_n c^2 \nonumber \\
&&  \left. - n_B t_{\rm FG}^\star(n_B,\delta_\beta) - n_B e_{\rm is}(n_B) \right] .\nonumber 
\end{eqnarray}
For each value of $n_B$, this equation can be numerically solved for $\delta_\beta$, as proposed by \citet{Essick2021}; the expression is slightly more involved because of the explicit inclusion of the kinetic term in Eq.~(\ref{eq:meta}), but the numerical complexity is the same as for Eq.~(22) in \cite{Essick2021}.

The solution of Eq.~(\ref{final}) requires knowledge of the isoscalar energy functional $e_{\rm is}(n_B)$.
A simple harmonic approximation \citep{Essick2021} is correct around saturation, but it is not sufficient (see e.g. Fig.~5 in \citet{Margueron2018}) if one wants to extract the behaviour of the functional at very low density, which is the domain better constrained by the ab initio calculations, and it is also very important for a correct calculation of the crust \citep{Dinh2021a}. 
We then used the meta-model labelled `ELFc' in \citet{Margueron2018}, truncated at order $N$.
It was shown that, when truncating the expansion at order $N=4$ (and even $N=3$), the meta-model gives a very good reproduction of realistic functionals at low density (see \citet{Margueron2018} and Sect.~\ref{sec:code-valid} for a discussion on the truncation order).
Defining $x=(n_B-n_{\rm sat})/(3n_{\rm sat})$, we write:
\begin{equation}
e_{\rm is}(n_B)=\sum_{k=0}^N \frac{v_k^{\rm is}}{k!}x^k u_k(x) \ ,
\label{eq:e0}
\end{equation}
where the function $u_k$,
\begin{equation}
u_k(x)=1-(-3x)^{N+1-k} e^{-b n_B/n_{\rm sat}}
\label{eq:uk}
\end{equation}
with $b=10\ln 2$ \citep{Margueron2018}, ensures the correct zero-density limit, and the parameters $v_i^{\rm is}$ are directly connected to the so-called isoscalar nuclear empirical parameters. 
At order $N=4$ we obtain the following:
\begin{eqnarray}
v_0^{\rm is}&=& E_{\rm sat} - t_{\rm FG, sat} (1+\kappa_{\rm sat}) \label{eq:v0} \;,\\
v_1^{\rm is}&=& - t_{\rm FG, sat} (2+5\kappa_{\rm sat}) \label{eq:v1} \;,\\
v_2^{\rm is}&=& K_{\rm sat} - 2t_{\rm FG, sat} (5\kappa_{\rm sat}-1) \label{eq:v2} \;,\\
v_3^{\rm is}&=& Q_{\rm sat} - 2t_{\rm FG, sat} (4-5\kappa_{\rm sat}) \label{eq:v3}\;, \\
v_4^{\rm is}&=& Z_{\rm sat} - 8t_{\rm FG, sat} (-7 + 5 \kappa_{\rm sat}) \label{eq:v4} \;. 
\end{eqnarray}
Once the isoscalar parameters at order $N$ ($E_{\rm sat}$, $K_{\rm sat}$, $Q_{\rm sat}$, $Z_{\rm sat}$ for $N=4$) are picked, Eq.~(\ref{final}) is solved for $N+1$ different sub-saturation density points, $n_{B,j}$, corresponding to $N+1$ points --- $x_j$, $j=1\dots N+1$ --- which have to be defined. 
The isovector empirical parameters ($E_{\rm sym},L_{\rm sym},K_{\rm sym},Q_{\rm sym}$, $Z_{\rm sym}$ for $N=4$) can thus be analytically obtained by matrix inversion \citep{Mondal2022}:
\begin{small}
\begin{equation}
\label{eq:sym-mat}
   \left |
   \begin{array}{c}
   E_{\rm sym}\\ L_{\rm sym}\\   K_{\rm sym}\\ Q_{\rm sym}\\ Z_{\rm sym}
   \end{array}
\right |=
\left |
            \begin{array}{ccccc} 
U_{01} &  x_1 U_{11}  & \frac{x_1^2}{2} U_{21} & \frac{x_1^3}{6}U_{31} & \frac{x_1^4}{24}U_{41} \\   
U_{02} &  x_2 U_{12}  & \frac{x_2^2}{2} U_{22} & \frac{x_2^3}{6}U_{32} & \frac{x_2^4}{24}U_{42} \\   
U_{03} &  x_3 U_{13}  & \frac{x_3^2}{2} U_{23} & \frac{x_3^3}{6}U_{33} & \frac{x_3^4}{24}U_{43} \\   
U_{04} &  x_4 U_{14}  & \frac{x_4^2}{2} U_{24} & \frac{x_4^3}{6}U_{34} & \frac{x_4^4}{24}U_{44} \\
U_{05} &  x_5 U_{15}  & \frac{x_5^2}{2} U_{25} & \frac{x_5^3}{6}U_{35} & \frac{x_5^4}{24}U_{45}
\end{array}  \right |^{-1} 
   \left |
   \begin{array}{c}
   e^1_{\rm iv}\\  e^2_{\rm iv} \\    e^3_{\rm iv}\\   e^4_{\rm iv}\\ e^5_{\rm iv}
   \end{array}
\right | \ ,
\end{equation}
\end{small}
with $U_{nk}=u_n(x_k)$ and
 \begin{eqnarray}
e^k_{\rm iv}&=&e_{\rm iv}(x_k) + \frac{5}{9}U_{0k} t_{\rm FG, sat} [1+(\kappa_{\rm sat}+3\kappa_{\rm sym})]  \nonumber \\ 
 &+&\frac{5}{9}x_k U_{1k} t_{\rm FG, sat} [2+5(\kappa_{\rm sat}+3\kappa_{\rm sym})]   \nonumber \\ 
&+&\frac{10}{9}\frac{x_k^2}{2} U_{2k} t_{\rm FG, sat} [-1+5(\kappa_{\rm sat}+3\kappa_{\rm sym})] \nonumber \\ 
&+&\frac{10}{9}\frac{x_k^3}{6}U_{3k} t_{\rm FG, sat} [4-5(\kappa_{\rm sat}+3\kappa_{\rm sym})] \nonumber     \\
&+&\frac{40}{9}  \frac{x_k^4}{24}U_{4k} t_{\rm FG, sat} [-7+5(\kappa_{\rm sat}+3\kappa_{\rm sym})] \ .
\label{eq:viv}
\end{eqnarray}
This `inversion procedure' thus allows one to extract the nucleonic functional $e_{\rm nuc}(n_B, \delta)$ (see Eqs.~\eqref{eq:enuc} and \eqref{eq:meta}) from a given beta-equilibrated EoS $\mathcal{E}_\beta(n_B)$ and a set of isoscalar parameters\footnote{Whenever the isoscalar empirical parameters are not provided, it is possible to assign specific values to them. In the numerical tool we propose (see Sect.~\ref{sec:results}), whenever the isoscalar empirical parameters are not provided in the original model, the latter are fixed to those of the BSk24 functional ($n_{\rm sat}=0.1578$~fm$^{-3}$, $E_{\rm sat}=-16.048$~MeV, and $K_{\rm sat}=245.5$~MeV; see \citet{Goriely2013}) in order to perform the inversion procedure at order $N=2$.}.
All isoscalar and isovector parameters being known, the crust EoS and the crust--core transition point are calculated from the crust code originally described in \citet{Carreau2019, Dinh2021b, Dinh2021c} (see Sect.~\ref{sec:crust}).

A first application of this method consists of assigning specific values of the isoscalar empirical parameter corresponding to those of given nuclear models, which allows one to reconstruct the isovector parameters when not known at the same order as the isoscalar ones; with this set of parameters, it is thus possible to compute a unified EoS.
This is of particular interest for EoSs that have only been computed for the core (e.g. EoSs based on microscopic approaches not yet available for the crust), thus allowing for a unified treatment of all the regions of the NS.
Results for this application are discussed in Sect.~\ref{sec:code-valid}.

A second and wider application of this approach is connected to Bayesian inference from GW data for which agnostic EoSs are often used. 
Indeed, for each agnostic realisation $\mathcal{E}_\beta(n_B)$, it is possible to sample the values of the (unknown) parameters $(n_{\rm sat}, E_{\rm sat}, K_{\rm  sat}, Q_{\rm sat},m_{\rm sat}^\star,\Delta m_{\rm sat}^\star)$, and eventually $Z_{\rm sat}$ for an order-4 reconstruction, from flat (or gaussian) distributions in the following range \citep{Dinh2021c}:
\begin{eqnarray}
n_{\rm sat}&=& 0.16 \pm 0.01\  \mathrm{fm}^{-3} \label{eq1} \\
E_{\rm sat}&=& -16.00 \pm 1.00\ \mathrm{MeV} \label{eq2} \\
K_{\rm sat}&=& 230 \pm 40 \ \mathrm{MeV} \label{eq3} \\
Q_{\rm sat}&=& 0.0 \pm 1000.0    \ \mathrm{MeV} \label{eq4} \\
Z_{\rm sat}&=& 0.0 \pm 3000.0 \ \mathrm{MeV} \\
m_{\rm sat}^\star/m&=& 0.7 \pm 0.1 \label{eq5} \\
\Delta m_{\rm sat}^\star/m &=&(m^\star_n-m^\star_p)/m = 0.1 \pm 0.1 \ .  \label{eq6}
\end{eqnarray}
Once the isoscalar parameters are picked, and for each sub-saturation density point $x_j$, Eq.~(\ref{esym}) gives $e_{\rm iv}(n_B)$ and, consequently, the neutron energy $e_{\rm nuc}(n_B,1)$ from Eq.~(\ref{eq:meta}).  
The model can thus be filtered through chiral effective field theory (EFT) calculations (see Sect.~\ref{sec:bayes}).
If retained, the isovector parameters can be calculated via Eq.~(\ref{eq:sym-mat}); thus, a consistent and unified crust--core EoS can be constructed.

\subsection{Crust equation of state}
\label{sec:crust}

The crust EoS is calculated as described in \citet{Carreau2019} (see also \citet{Dinh2021a, Dinh2021b, Dinh2021c}) within a compressible liquid-drop (CLD) model approach.
Although not as microscopic as a full density functional treatment, this method was recently shown to provide results in good agreement with extended Thomas-Fermi calculations both at zero \citep{Grams2022} and at finite temperature \citep{Carreau2020} (see also Fig.~2 in \citet{Carreau2019} for an illustration of the performance of the CLD model we used to reproduce experimental masses).
Moreover, it makes it possible to quantitatively estimate the model dependence of the results by comparing different EoS models at a reduced computational cost.
We only recall the main points of this method here for completeness.
Because of the zero-temperature approximation, we employ the one-component plasma approach, that is, we consider one nuclear cluster at each (increasing) density in the crust for the composition.
At each layer of the crust, characterised by a pressure $P$ and baryon density $n_B$, matter is modelled as a periodic lattice consisting of Wigner-Seitz cells of volume $V_{\rm WS}$ containing one type of cluster with $Z$ protons of mass $m_p$ and $A-Z$ neutrons of mass $m_n$ ($A$ being the cluster total mass number), immersed in a uniform gas of electrons and, in the inner crust, of neutrons too. 
For a given thermodynamic condition, defined by the baryonic density $n_B$, the total energy density of the inhomogeneous system can thus be written as
\begin{equation}
    \mathcal{E}_{\rm inhom} = \mathcal{E}_e +  \mathcal{E}_{\rm g} (1-u) + \frac{E_i}{V_{\rm WS}} \ ,
    \label{eq:energy-density}
\end{equation}
where $\mathcal{E}_e(n_e)$ is the electron gas energy density, $\mathcal{E}_{\rm g} = \mathcal{E}_{\rm nuc}(n_{\rm g}, \delta_{\rm g}) $ is the neutron-gas energy density (including the rest masses of nucleons) at baryonic density $n_{\rm g} = n_{{\rm g}n}$ and isospin asymmetry $\delta_{\rm g} = 1$ (no protons are present in the gas), and $u=\frac{A/n_i}{V_{\rm WS}}$ is the volume fraction of the cluster with internal density $n_i$. Finally, the third term on the right hand side of Eq.~(\ref{eq:energy-density}), $- \mathcal{E}_{\rm g} u$, accounts for the interaction between the cluster and the neutron gas, that we treated in the excluded-volume approach.
In the outer crust, $\mathcal{E}_{\rm g}$ is set to zero.
The last term in Eq.~(\ref{eq:energy-density}) is the cluster energy per Wigner-Seitz cell: 
\begin{equation}
    E_i = M_i c^2 + E_{\rm bulk} + E_{\rm Coul} + E_{\rm surf +  curv} \ ,
    \label{eq:Ei}
\end{equation}
where $M_i = (A-Z)m_n + Zm_p$ is the total bare mass of the cluster, $E_{\rm bulk} = A e_{\rm nuc}(n_i, 1-2Z/A)$ is the cluster bulk energy, and $E_{\rm Coul} + E_{\rm surf + curv} = V_{\rm WS} (\mathcal{E}_{\rm Coul} + \mathcal{E}_{\rm surf + curv})$ accounts for the total interface energy, that is, the Coulomb interaction between the nucleus and the electron gas, as well as the residual interface interaction between the nucleus and the surrounding dilute nuclear-matter medium.
For the bulk energy density of the ions --- $\mathcal{E}_{\rm bulk} = e_{\rm nuc}n_i$ --- and that of the neutron gas, we use the same functional expression, that is, the meta-model approach of \citet{Margueron2018} (see Eq.~(\ref{eq:edens-meta})).
We consider here only spherical clusters, since the so-called pasta phases that may appear at the bottom of the inner crust are expected to have only a small impact on the EoS; therefore, the Coulomb energy density is given by 
\begin{equation}
    \mathcal{E}_{\rm Coul}  = \frac{2}{5}\pi (e n_i r_N)^2 u \left(\frac{1-I}{2}\right)^2 \left[ u+ 2  \left( 1- \frac{3}{2}u^{1/3} \right) \right] \ , 
    \label{eq:Fcoul}
\end{equation}
with $e$ being the elementary charge, $I = 1-2Z/A$, and $r_N = (3A/(4 \pi n_i))^{1/3}$.
For the surface and curvature contributions, we employed the same expression as in \citet{Maruyama2005} and \citet{Newton2013}, that is
\begin{equation}
\mathcal{E}_{\rm {surf}} + \mathcal{E}_{\rm {curv}} =\frac{3u}{r_N} \left( \sigma_{\rm s}(I) +\frac{2\sigma_{\rm c}(I)}{r_N} \right) \ , 
\label{eq:interface}   
\end{equation}
where $\sigma_{\rm s}$ and $\sigma_{\rm c}$ are the surface and curvature tensions \citep{Ravenhall1983}, and
\begin{eqnarray}
\sigma_{\rm s}(I) &=& \sigma_0 \frac{2^{p+1} + b_{\rm s}}{y_p^{-p} + b_{\rm s} + (1-y_p)^{-p}} \ , \\
\sigma_{\rm c} (I) &=& 5.5 \sigma_{\rm s}(I) \frac{\sigma_{0, {\rm c}}}{\sigma_0} (\beta -y_p) \ ,
\label{eq:sigma0}
\end{eqnarray}
where $y_p = (1-I)/2$ and the surface parameters $(\sigma_0, \sigma_{0, {\rm c}}, b_{\rm s}, \beta)$ were optimised for each set of bulk parameters and effective mass to reproduce the experimental nuclear masses in the 2020 Atomic Mass Evaluation (AME) table \citep{ame2020}, while we set $p=3$.
The EoS and composition of the crust was then obtained by variationally minimising the energy density of the Wigner-Seitz cell with $(A, I, n_i, n_e, n_{\rm g})$ as variational variables, under the constraint of baryon number conservation and charge neutrality holding in every cell \citep{Carreau2019, Dinh2021a, Dinh2021b}.

\subsection{Global neutron-star properties}
\label{sec:NS-prop}

Once the EoSs $P(\mathcal{E})$ are generated, the properties of a static NS are calculated from the Tolman-Oppenheimer-Volkoff (TOV) equations \citep{Tolman1939, Oppenheimer1939}: 
\begin{eqnarray}
    \label{eq:TOV1}
    \frac{{\rm d}P(r)}{{\rm d}r} &=& -\frac{G \mathcal{E}(r)\mathcal{M}(r)}{c^2 r^2}
    \biggl[1+\frac{P(r)}{\mathcal{E}(r)}\biggr] \nonumber \\ 
    && \times \biggl[1+\frac{4\pi P(r)r^3}{c^2\mathcal{M}(r)}\biggr]\biggl[1-\frac{2G\mathcal{M}(r)}{c^2 r}\biggr]^{-1} \ ,
    \end{eqnarray}
with $\mathcal{E}$ being the total energy density of the (beta-equilibrated) system and
\begin{equation}
    \label{eq:TOV2}
    \mathcal{M}(r) = \frac{4\pi}{c^2} \int_0^r 
    \mathcal{E}(r')r'^2{\rm d}r' \ .
\end{equation}
The integration of Eqs.~\eqref{eq:TOV1}-\eqref{eq:TOV2} until the surface of the star defines its radius $R$ and mass $M=\mathcal{M}(r=R)$.
In addition, the dimensionless tidal deformability $\Lambda = \lambda(r=R)$ is calculated as
\begin{equation}
  \lambda(r) = \frac{2}{3} k_2(r) \left[\frac{rc^2}{G\mathcal{M}(r)}\right]^5 \ , 
\label{eq:lambda} 
\end{equation} 
with the tidal Love number function $k_2(r)$ given by~\citep{Hinderer2008, Hinderer2010}
\begin{align}
    k_2 =& \,\frac{8C^5}{5}(1-2C)^2\big[2+2C(y-1)-y\big]\nonumber \\
    &\times \Big\{2C\big[6-3y+3C(5y-8)\big] \nonumber \\ 
    &+ 4C^3\big[13-11y+C(3y-2)+2C^2(1+y)\big] \nonumber \\ 
    &+ 3(1-2C)^2\big[2-y+2C(y-1)\big] \ln (1-2C) \Big\}^{-1} \ ,
    \label{eq:k2}
\end{align}
where we introduce the dimensionless compactness parameter
\begin{equation}
    C(r) = \dfrac{G\, \mathcal{M}(r)}{r\, c^2}\, ,
\end{equation}
and the function $y(r)$ is obtained by integrating the following differential equation with the boundary condition $y(0)=2$: 
\begin{equation}
    \label{y_eq}
    r \dfrac{dy}{dr}+ y(r)^2 + F(r) y(r) + Q(r)=0 \, ,
\end{equation}
\begin{equation}
    \label{F(r)}
    F(r)=\frac{1-4\pi G r^2(\mathcal{E}(r)-P(r))/ c^4}{1-2G\mathcal{M}(r)/(r c^2)}\, ,
\end{equation}
\begin{align}
    \label{Q(r)}
    Q(r)=&\frac{4 \pi G r^2/c^4}{1-2G\mathcal{M}(r)/(r c^2)}\Biggl[5\mathcal{E}(r)+9P(r)\nonumber \\
    &+\frac{\mathcal{E}(r)+P(r)}{c_s(r)^2} c^2-\frac{6\,  c^4}{4\pi r^2 G}\Biggr] \nonumber \\
    &-4\Biggl[ \frac{G(\mathcal{M}(r)/(r c^2)+4\pi r^2 P(r)/c^4)}{1-2G\mathcal{M}(r)/(r c^2)}\Biggr]^2\, ,
\end{align}
with $ c_s=c \sqrt{dP/d\mathcal{E}}$ denoting the sound speed.

We also recall that the effective tidal deformability $\tilde{\Lambda}$ of the binary, which can be extracted from GW measurements, is expressed in the form of the component masses, $m_2 \le m_1$, and the two corresponding dimensionless tidal deformabilities, $\Lambda_1 = \lambda_1(R_1)$ and $\Lambda_2 = \lambda_2(R_2)$, as
\begin{equation}
    \tilde{\Lambda} = \frac{16}{13} \frac{(m_1+12m_2)m_1^4 \Lambda_1 + (m_2+12m_1)m_2^4 \Lambda_2}{(m_1+m_2)^5} \ ,
    \label{eq:lambdatil}
\end{equation}
while the chirp mass associated with a binary system reads
\begin{equation}
    \mathcal{M}_{\rm c} = \frac{(m_1 m_2)^{3/5}}{(m_1 + m_2)^{1/5}} = \frac{q^{3/5} m_1}{(1+q)^{1/5}} \ ,
    \label{eq:mchirp}
\end{equation}
with $q = m_2/m_1$ being the mass ratio.

\section{Numerical results}
\label{sec:results}

In this section, we present the results obtained using the Crust (Unified) Tool for Equation-of-state Reconstruction (CUTER) code\footnote{The first release of the code is currently available for the LIGO-Virgo-KAGRA collaboration members and is publicly available at Zenodo (version labelled as `Version v1'): \url{https://zenodo.org/records/10781539} (DOI: 10.5281/zenodo.10781539).},
based on the method described in Sect.~\ref{sec:method}.
We point out that the reconstructed EoS is only valid until around saturation density, where the density points $n_{B,j}$ are chosen.
The final EoS is then built by matching the reconstructed EoS to the original one either at the crust--core transition or until the first point in the original (high-density) EoS table, if the latter is after the crust--core transition.
However, since the nuclear empirical parameters used for the crust are consistent with those used for the core EoS around saturation density, a continuous and smooth behaviour in the energy density is ensured.

\subsection{Code validation}
\label{sec:code-valid}

In order to validate the procedure, we start by considering known beta-equilibrated EoSs, namely the PCP(BSk24) \citep{Pearson2018, Pearson2018err}, the RG(SLy4) \citep{Gulminelli2015}, the GPPVA(DDME2) \citep{Grill2012, Grill2014}, and the GPPVA(TW) \citep{Grill2014} ones, available on the \texttt{CompOSE} database\footnote{\url{https://compose.obspm.fr}} \citep{Typel2015, Typel2022}. 
They are based on the BSk24 \citep{Goriely2013}, the SLy4 \citep{Chabanat1998}, the DDME2~\citep{Lalazissis2005}, and the TW99 \citep{Typel1999} energy-density functionals, respectively.
The choice of these models was driven by the fact that they span different values of the nuclear-matter parameters, which are consistent with current knowledge of nuclear physics, and they all yield EoSs that are able to support $2 M_\odot$ NSs.
Finally, the corresponding nuclear empirical parameters up to order $N=4$ (see Table~\ref{tab:param}) are available for these functionals \citep{Margueron2018, Dinh2021a}, so different orders of the reconstruction can be tested.
Although we only show a limited number of models here, as illustrative examples, we performed code validation and automated tests on a larger set of available EoSs.

We start by fixing the isoscalar parameters as those characterising the corresponding functional\footnote{The nuclear empirical parameters for each EoS are also available in \texttt{JSON} format on the \texttt{CompOSE} database. For the higher order parameters, not available in the \texttt{JSON} meta-data and needed for the reconstruction at higher orders, we took the parameter values listed in Table~XI of \citet{Margueron2018} and Table~1 in \citet{Dinh2021a}. As for the effective mass and isospin splitting, if not available in the meta-data, they are set to 1 and 0, respectively, except for BSk24, where the known values of the effective mass $m^\star_{\rm sat}/m = 0.8$ and effective-mass splitting $\Delta m^\star_{\rm sat}/m = 0.2$ are taken from \citet{Goriely2013}.} and $N$ points in density, $n_{B,j}$, around saturation ($N$ being the order of the reconstruction; see Table~\ref{tab:density-points}).
The isovector parameters are then calculated using the inversion procedure described in Sect.~\ref{sec:method} (see Eq.~(\ref{eq:sym-mat})).
We emphasise that the obtained parameter values, and thus the EoS reconstruction, may depend on the choice of the density points $n_{B,j}$.
However, the impact on NS macroscopic properties such as the mass--radius relation remains small, as long as the original (input) EoS covers a baryon density range that does not require considerable extrapolation to obtain the EoS at the chosen $n_{B,j}$ required for the reconstruction at a given order $N$.

\begin{table}[!ht]
    \centering
    \caption{Baryon number density points $n_{B,j}$ (in fm$^{-3}$) for the EoS reconstruction according to different orders $N$.}
    \begin{tabular}{c|c|c|c|c|c}
        $N$ &   $n_{B,1}$  &  $n_{B,2}$ &  $n_{B,3}$ &  $n_{B,4}$ &  $n_{B,5}$ \\
        \hline
         2 & 0.12 & 0.14 & 0.16 & - & - \\
         3 & 0.10 & 0.12 & 0.14 & 0.16 & - \\
         4 & 0.08 & 0.10 & 0.12 & 0.14 & 0.16\\
    \end{tabular}
    \label{tab:density-points}
\end{table}

To evaluate the quality of the reconstruction, we start by listing the nuclear empirical parameters for the four considered models in Table~\ref{tab:param}: the first line corresponds to the parameter set of the original model, while the following lines indicate the (input) isoscalar parameters and the reconstructed isovector ones for different orders of the reconstruction.
We can see that the inversion procedure yields a good reproduction of the low-order parameters, while larger deviations are observed for the high-order ones.
This can be understood from the fact that the reconstruction requires us to fix a number of given points in density (see Table~\ref{tab:density-points}), chosen around and below saturation.
In this density regime, the contribution of the high-order terms is almost negligible; thus, the discrepancies in the high-order terms arise from the capability of the polynomial expansion to reproduce the functional behaviour (thus to recover the higher order parameters) when extrapolated far below or above saturation.
However, the NS mass--radius relation is only slightly impacted since it is dominated by the high-density part, where in any case the original EoS is matched.

\begin{table*}[!ht]
    \centering
    \caption{Nuclear empirical parameters for the BSk24, SLy4, DDME2, and TW99 models used in this work. }
    \begin{tabular}{c|c|c|c|c|c||c|c|c|c|c}
           &   $n_{\rm sat}$  &  $E_{\rm sat}$ & $K_{\rm sat}$ &  $Q_{\rm sat}$ &  $Z_{\rm sat}$
          & $E_{\rm sym}$ & $L_{\rm sym}$ & $K_{\rm sym}$ & $Q_{\rm sym}$ & $Z_{\rm sym}$ \\
        \hline
         BSk24 set & 0.1578 & -16.048 & 245.5 & -274.5 & 1184.2 &  30.0 & 46.4 & -37.6 & 710.9 & -4031.3   \\
         order 2 & 0.1578 & -16.048 & 245.5 & - & - & 29.9 & 45.1 & -55.2 & - & - \\ 
         order 3 & 0.1578 & -16.048 & 245.5 & -274.5 & - & 29.9 & 45.1 & -55.7 & 573.3 & -  \\
         order 4 & 0.1578 & -16.048 & 245.5 & -274.5 & 1184.2 & 29.9 & 45.1 & -62.5 & 280.1 & -8326.2  \\
         \hline \hline 
         
         SLy4 set & 0.159 & -15.97 & 230.0 & -363.11 & 1587.0 & 32.0  & 46.0 & -119.7 & 521.0 & -3197.0   \\
         order 2 & 0.159 & -15.97 & 230.0 & - & - & 31.8 & 44.5 & -122.4 & - & - \\
         order 3 & 0.159 & -15.97 & 230.0 & -363.11 & - & 31.8 & 44.3 & -139.5 & 310.6 & - \\
         order 4 & 0.159 & -15.97 & 230.0 & -363.11 & 1587.0 & 31.8 & 44.3 & -139.3 & 319.5 & -3789.2  \\
        \hline \hline 
        
         DDME2 set & 0.152 & -16.14 & 251.0 & 479.0 & 4448.0 &  32.3 & 51.0 & -87.2 & 777.0 & -7048.0   \\
         order 2 & 0.152 & -16.14 & 251.0 & - & - & 32.5 & 51.0 & -128.3 & - & - \\
         order 3 & 0.152 & -16.14 & 251.0 & 479.0  & - & 32.5 & 51.0 & -107.9 & 496.2 & - \\
         order 4 & 0.152 & -16.14 & 251.0 & 479.0 & 4448.0 & 32.5 & 51.0 & -97.1 & 1189.9 & 7237.8  \\
         \hline \hline 
        
         TW99 set & 0.153 & -16.25 & 240.0 & -540.0 & 3749.0 & 32.8 & 55.0 & -125.0 & 539.0 & -3307.0  \\
         order 2 & 0.153 & -16.25 & 240.0 & - & - & 33.0 & 54.8 & -128.2 & - & - \\
         order 3 & 0.153 & -16.25 & 240.0 & -540.0 & - & 33.0 & 54.7 & -147.9 & 256.5 & - \\
         order 4 & 0.153 & -16.25 & 240.0 & -540.0 & 3749.0 & 33.0 & 54.7 & -141.7 & 626.6 & 996.1  \\
    \end{tabular}
     \tablefoot{The saturation density $n_{\rm sat}$ is given in units of fm$^{-3}$, while the other nuclear empirical parameters are in units of MeV.}
    \label{tab:param}
\end{table*}

\begin{figure}
    \begin{center}
    \includegraphics[width=0.5\textwidth]{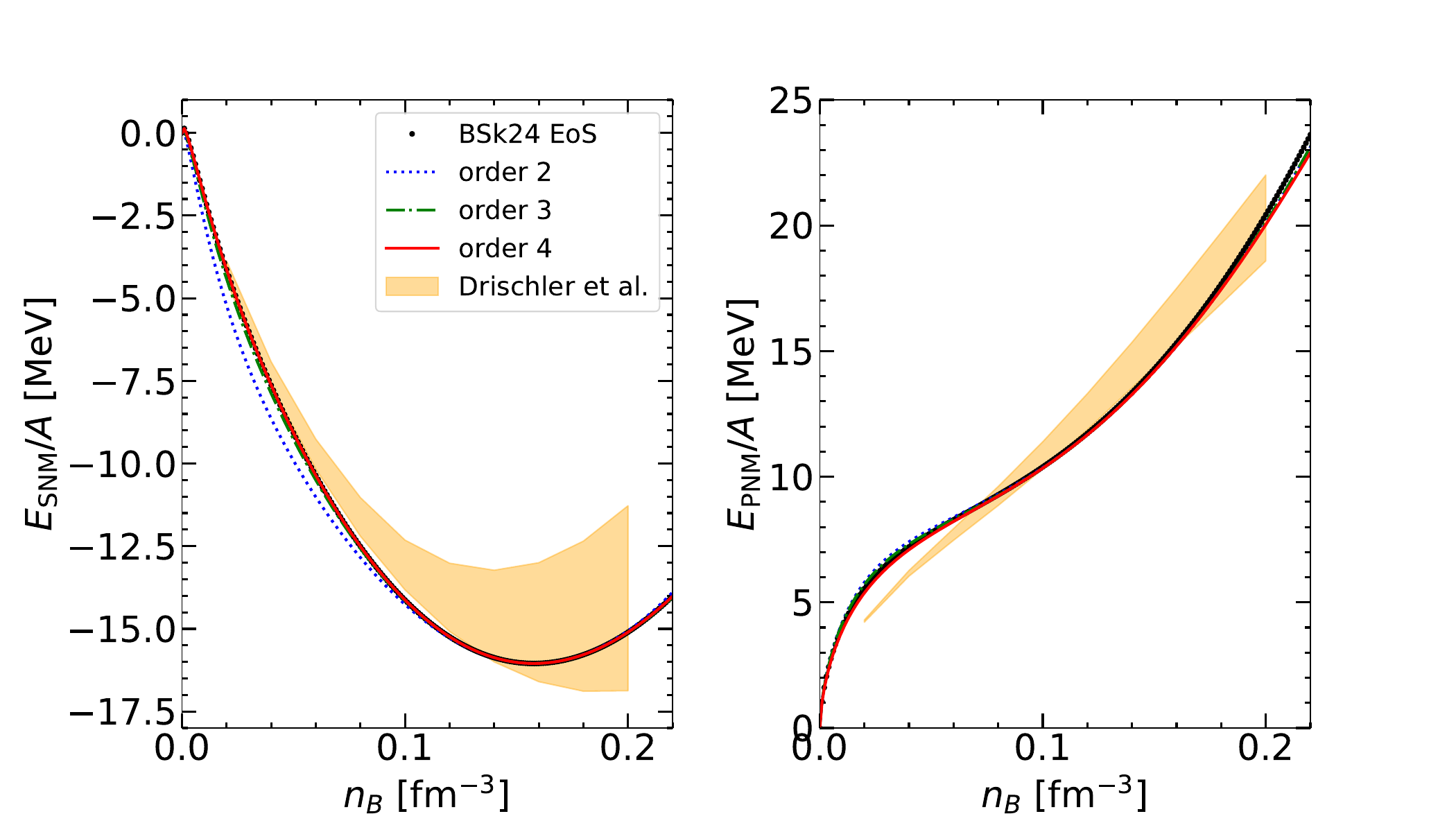}
    \includegraphics[width=0.5\textwidth]{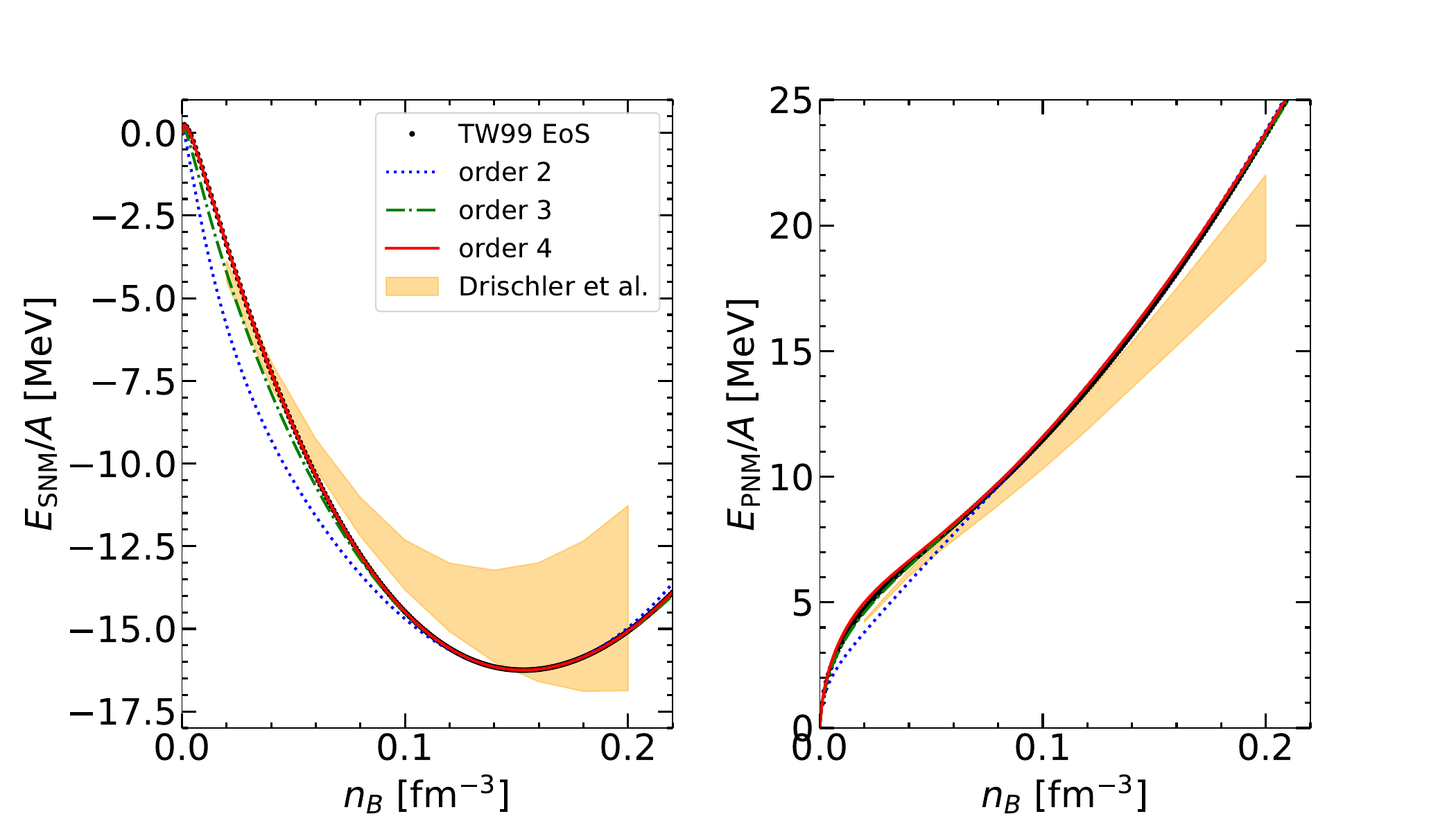}
    \end{center}
    \caption{Energy per baryon versus baryon number density for symmetric nuclear matter (SNM; left panels) and for pure neutron matter (PNM; right panels), for the BSk24 (top panels) and the TW99 (bottom panels) models. Dots correspond to the results obtained using the original set of isoscalar and isovector parameters, while dotted blue lines, dash-dotted green lines, and solid red lines correspond to the results from the reconstruction at order 2, 3, and 4, respectively. For comparison, the chiral-EFT calculations of \citet{Drischler2016} are also shown (yellow bands; see text for details).
    }
    \label{fig:comp-order-hm}
\end{figure}

With the complete set of parameters, we first calculate the EoS of pure neutron matter and symmetric matter.
The correct reproduction of the homogeneous matter is particularly important when performing Bayesian analyses since the chiral-EFT filter, which indeed operates on homogeneous matter, can be applied to discriminate models.
In Fig.~\ref{fig:comp-order-hm}, we show the energy per baryon of symmetric nuclear matter (left panels) and pure neutron matter (right panels) for the BSk24 (top panels) and TW99 (bottom panels) models as illustrative examples; the dots are obtained with the full parameter set of the original model, while the dotted, dash-dotted, and solid curves correspond to the results obtained with the reconstruction at order 2, 3, and 4, respectively.
We can see that for the symmetric sector (left panels) the reconstruction is very close to the original model, with a slight deviation at very low density for the reconstructions at order 2. 
This is somehow expected since only the isoscalar parameters enter in the determination of the symmetric-matter energy (see Eqs.~(\ref{eq:meta}), (\ref{eq:v0})-(\ref{eq:v4})), and these are fixed as input.
The small deviation observed can therefore be attributed to the capability of the meta-model truncated at a given order to reproduce the original functional (see also the discussion in \citet{Margueron2018}).
On the other hand, in the calculation of the pure neutron-matter energy, the isovector parameters, which are calculated through the inversion procedure in Eq.~(\ref{eq:sym-mat}), play an important role.
Since the reconstruction gives values of the isovector parameters close to those of the original model but not exactly identical, deviations can appear far from saturation density (i.e. away from the chosen $n_{B,j}$ points in density; see Table~\ref{tab:density-points}).
This is observed, for the considered cases, for the order-2 reconstruction for TW99 at very low density, where high-order parameters can also be influential (see lower right panel of Fig.~\ref{fig:comp-order-hm}), while the inversion procedure yields a very good reconstruction at any order for BSk24 (upper right panel of Fig.~\ref{fig:comp-order-hm}).
For comparison, we also show the band of the chiral-EFT calculations of \citet{Drischler2016} (orange shaded area). 
It is interesting to see that the reproduction of the original model is, in both of the shown cases (and for all EoS models that we tested), very precise at order 4 in spite of the relatively poor reconstruction of the isovector nuclear-matter parameters shown in Table~\ref{tab:param}. 
This underlines the fact that significant degeneracy exists in the nuclear-empirical-parameter space, and caution should be taken when attributing specific physical effects to individual parameters, particularly of order 2 or higher.
We also note that the quality of the reconstruction depends on the original input EoS as well (density grid, numerical precision, etc.), and thus on the numerical computation of the EoS.

\begin{figure*}
    \begin{center}
    \includegraphics[scale=0.3]{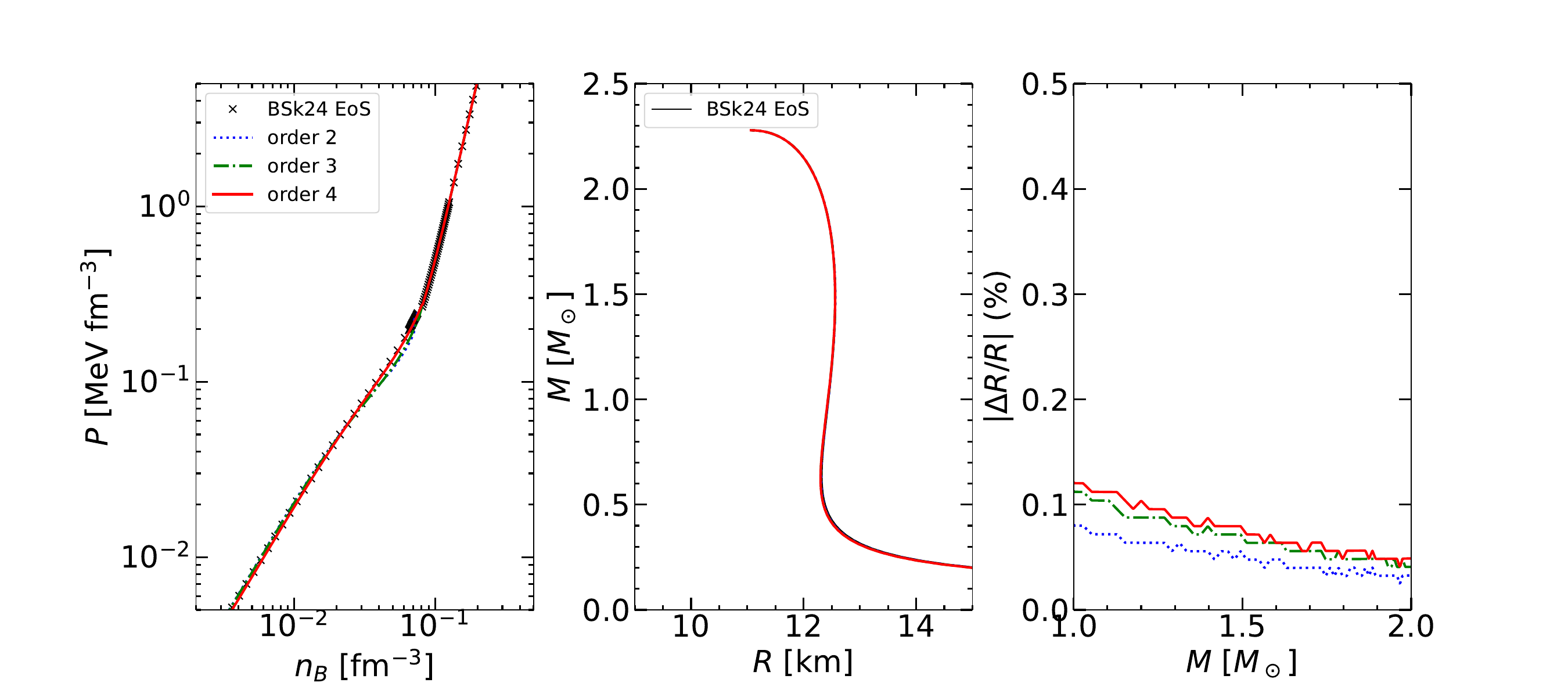}
    \includegraphics[scale=0.3]{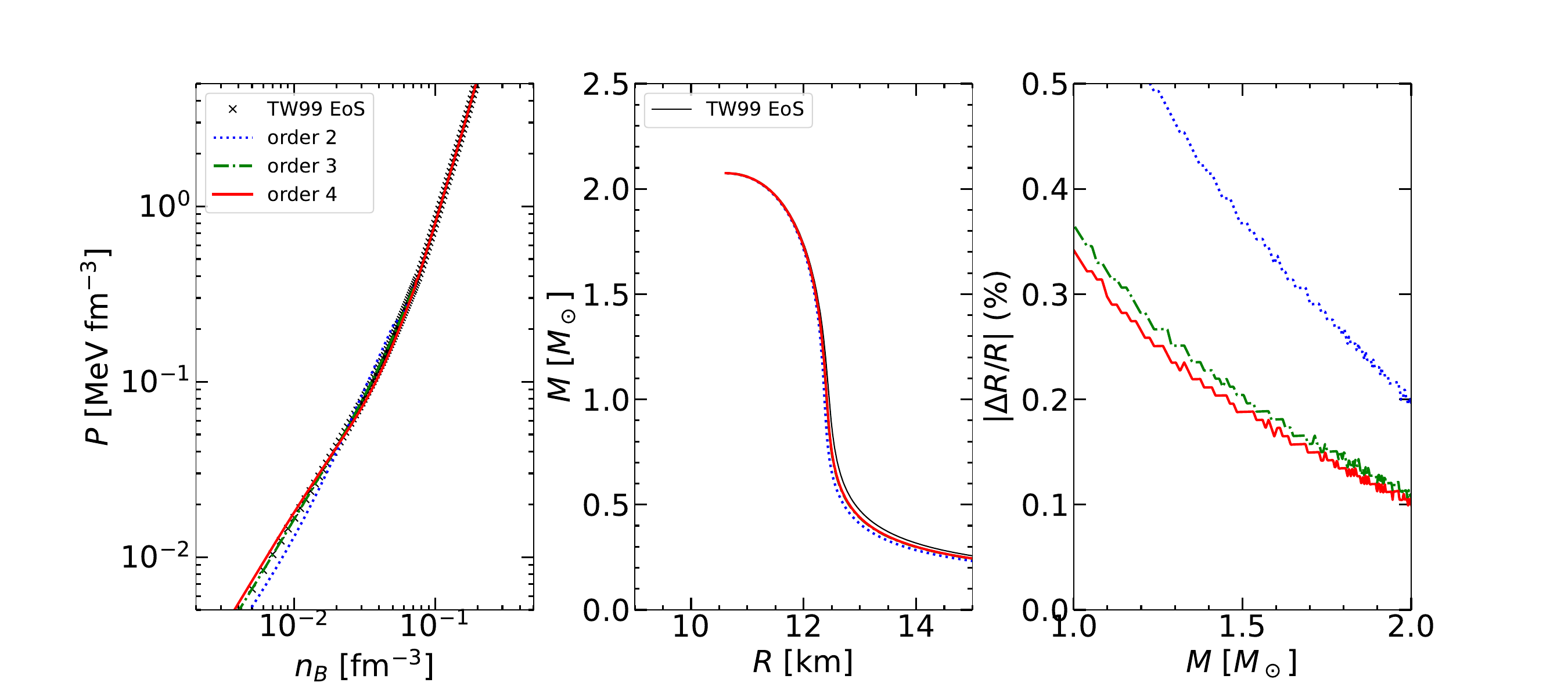}
    \end{center}
    \caption{Pressure versus baryon number density (left panels), mass-radius relation (middle panels), and absolute relative error on the radius (right panels) for different models. Crosses and solid black lines (left and middle panel, respectively) correspond to the original EoS, while dotted blue lines, dash-dotted green lines, and solid red lines correspond to the reconstruction at order 2, 3, and 4, respectively (see text for details).
    }
    \label{fig:comp-order}
\end{figure*}

We now discuss the beta-equilibrium EoS calculated with the parameter sets reconstructed at different orders and its impact on (global) NS properties. 
In Fig.~\ref{fig:comp-order} we display the EoS (left panels), the resulting mass-radius
relation (middle panels), and the relative absolute error on the radius (right
panels) for the BSk24 (top panels) and TW99 (bottom panels) models with respect to that obtained using the original EoS tables.
We can see that the EoS is reconstructed with very good accuracy, and the mass-radius curves are almost indistinguishable.
Instead, it might be surprising that the order-2 EoS, which generally does not yield the most accurate reconstruction, gives instead the lowest error on the radius in some cases, but this might be also due to the numerical interpolation of the EoS table in the TOV solver used.
In any case, the relative error on the radius remains very small: $\lesssim 0.5\%$ in all cases considered here.

Of particular importance for the NS modelling, and specifically for observables related to the crust physics, is the crust--core transition.
In Table~\ref{tab:cc}, we list the crust--core transition density and pressure for the four considered models, as indicated in the `original' EoS tables \citep{Pearson2018, Douchin01, Grill2014} and as predicted by the EoSs reconstructed at different orders.
We can see that the reconstruction procedure also gives very good predictions for the crust--core transition, the order-4 (order-2) EoS generally being the one yielding the better (worse) agreement with the values obtained in the original models.
These results can be deemed as satisfactory, especially considering that the exact value of crust--core transition pressure and density are known to depend not only on the EoS, but also on the many-body method used to determine it.
An example of this latter dependence can be seen by comparing the RG(SLy4) \citep{Gulminelli2015} and the DH(SLy4) \citep{Douchin01} EoSs, both based on the SLy4 functional; indeed, the crust--core transition density and pressure predicted by the RG(SLy4) EoS are $n_{\rm cc}=0.052$~fm$^{-3}$ and $P_{\rm cc}=0.16$~MeV~fm$^{-3}$, respectively, to be compared with the values reported in \citet{Douchin01}, namely, $n_{\rm cc}=0.076$~fm$^{-3}$ and $P_{\rm cc}=0.34$~MeV~fm$^{-3}$.

\begin{table}[!ht]
   \caption{Crust--core transition (baryon number density $n_{\rm cc}$ and pressure $P_{\rm cc}$) for the EoSs based on the BSk24, SLy4, DDME2, and TW99 models. 
   }
    \centering
    \begin{tabular}{c|c|c}
         & $n_{\rm cc}$ [fm$^{-3}$] & $P_{\rm cc}$ [MeV fm$^{-3}$] \\
         \hline
       PCP(BSk24) EoS  & 0.081 & 0.27 \\
       \hline
       order 2 & 0.071 & 0.19 \\
       order 3 & 0.076 & 0.23 \\
       order 4 & 0.080 & 0.27 \\
       \hline \hline
       DH(SLy4) EoS  & 0.076 & 0.34 \\
       \hline
       order 2 & 0.063 & 0.26 \\
       order 3 & 0.075 & 0.32 \\
       order 4 & 0.079 & 0.35 \\
       \hline \hline
       GPPVA(DDME2) EoS  & 0.072 & 0.41 \\
       \hline
       order 2 & 0.063 & 0.28 \\
       order 3 & 0.048 & 0.22 \\
       order 4 & 0.056 & 0.25 \\
       \hline \hline
       GPPVA(TW) EoS  & 0.074 & 0.38 \\
       \hline
       order 2 & 0.052 & 0.22 \\
       order 3 & 0.071 & 0.35 \\
       order 4 & 0.073 & 0.35 \\
    \end{tabular}
    \tablefoot{For each model, values in the first row correspond to those predicted by the original EoS tables (\citet{Pearson2018, Douchin01, Grill2014}), while values in the following rows correspond to those predicted by the EoS reconstructed at different orders.}
    \label{tab:cc}
\end{table}

These results show that the inversion procedure can be applied to different models for the EoS, thus allowing the reconstruction of the crust and core EoS in a consistent and unified way.
In order to evaluate the impact on the global NS properties of using such unified EoSs against a unique crust model, we performed a Bayesian analysis, as described in the next section.

\subsection{Bayesian analysis}
\label{sec:bayes}

To assess the uncertainties associated with the crust EoS, we performed a Bayesian analysis.
The nuclear meta-modelling is run at order $N=4$ up to a given baryon density $n_{\rm match}$, and the nuclear empirical parameters $\{ n_{\rm sat}, E_{\rm sat,sym}, L_{\rm sym}, K_{\rm sat,sym}, Q_{\rm sat,sym}, Z_{\rm sat,sym} \}$, together with the nucleon effective mass $m_{\rm sat}^\star/m$ and effective mass isosplit $\Delta m_{\rm sat}^\star/m,$ are sampled from a uniform non-informative prior (see Table~\ref{tab:nuc-emp-param}), as was done, for example, in \citet{Dinh2021c}.
In addition, for the present analysis, the $b$ parameter governing the low-density limit (see Eq.~(\ref{eq:uk})) is also varied in the $[1,10]$ range, and the surface parameter $p$ governing the isospin behaviour of the surface tension is fixed at $p=3$, as in \citet{Dinh2021c} (see e.g. \citet{Carreau2019} for a discussion).
The complete set of 13 parameters is denoted $\mathbf{X}_{\rm nuc}$.

To account for the current limited knowledge of the EoS and possible phase transitions in the NS core, as well as to illustrate the applicability of our procedure if an agnostic beta-equilibrated EoS is used at high density, we match the meta-model crust (and part of the outer core if $n_{\rm match}$ is above the crust--core transition) to an EoS built upon five  piecewise polytropes, each of which has the form $P = K \rho_B^\Gamma$, where $\rho_B = m_n n_B$ is the mass density and $K$ is a constant chosen to ensure continuity between the different EoS segments.
The transition densities among polytropes are randomly chosen between $n_{\rm match}$ and ten times saturation density, and each polytropic index $\Gamma$ is randomly chosen between 0 and 8; we denote the polytrope parameters (matching densities and $\Gamma$) as $\mathbf{X}_{\rm poly}$.
The piecewise polytropic EoS is calculated until the causality is reached.
Unlike other works in the literature (e.g. \citet{Hebeler2013, Raithel2023b}), we do not impose any condition on the spacing of the transition densities or the $\Gamma$ in order to allow for a wide variation of the EoS; therefore, the number of EoS segments may be less than five if causality is broken before the subsequent transition density is reached.
In the literature, five piecewise polytropes have been used to represent the (core) EoS matched with a unique crust in \citet{Raithel2023} and \citet{Raithel2023b}, while three polytropes have been employed, for example, in \citet{Read09} and \citet{Hebeler2013}.

The value of the matching density $n_{\rm match}$ at which the polytropes are glued is somehow arbitrary, and different values can be chosen.
However, we think that a value below the saturation density would be quite unrealistic, since it would allow the possibility of the occurrence of phase transitions below saturation, which actually seems to be excluded below about $2 n_{\rm sat}$ (see e.g. \citet{Tang2021, Christian2020}).
For this reason, we run the calculations for two values of the matching density; for illustrative purposes: $n_{\rm match} = 0.16$~fm$^{-3}$ and $n_{\rm match} = 0.32$~fm$^{-3}$.

\begin{table}[!ht]
    \caption{Minimum and maximum values of the parameter set $\mathbf{X}_{\rm nuc}$.}
    \centering
    \begin{tabular}{c|c|c}
         & $X_{\rm min}$ & $X_{\rm max}$ \\
         \hline
       $n_{\rm sat}$  & 0.15 & 0.17 \\
       $E_{\rm sat}$  & -17.0 & -15.0 \\
       $K_{\rm sat}$  & 190.0 & 270.0 \\
       $Q_{\rm sat}$  & -1000.0 & 1000.0 \\
       $Z_{\rm sat}$  & -3000.0 & 3000.0 \\
       $E_{\rm sym}$  & 26.0 & 38.0 \\
       $L_{\rm sym}$  & 10.0 & 80.0 \\
       $K_{\rm sym}$  & -400.0 & 200.0 \\
       $Q_{\rm sym}$  & -2000.0 & 2000.0 \\
       $Z_{\rm sym}$  & -5000.0 & 5000.0 \\
       $m^\star/m$  & 0.6 & 0.8 \\
       $\Delta m^\star/m$  & 0.0 & 0.2 \\
       $b$ & 1 & 10 
    \end{tabular}
    \label{tab:nuc-emp-param}
\end{table}

The posterior distributions of different observables $\mathcal{Y}$ under the set of constraints $\mathbf{c}$ are then obtained by marginalizing over the EoS parameters as
\begin{equation}
    P(\mathcal{Y}|\textbf{c}) = \prod_{k=1}^{N_{\rm par}} \int_{X_k^{\rm min}}^{X_k^{\rm max}} dX_k P(\mathbf{X}|\mathbf{c}) \delta (\mathcal{Y} - \mathcal{Y}(\mathbf{X})) \ ,
    \label{eq:bayes-postobs}
\end{equation}
where $N_{\rm par}$ is the number of parameters in the model, that is, the nuclear empirical parameters plus the polytrope parameters $\mathbf{X} = \{\mathbf{X}_{\rm nuc}; \mathbf{X}_{\rm poly} \}$.
The posterior distribution of the parameters conditioned by likelihood models of the different observations and constraints $\mathbf{c}$ is given by
\begin{equation}
    P(\mathbf{X}|\textbf{c}) = \mathcal{N} P(\mathbf{X}) \prod_k P(c_k|\mathbf{X}) \ ,
    \label{eq:bayes-postpar}
\end{equation}
where $P(\mathbf{X})$ is the prior and $\mathcal{N}$ is a normalization factor. 
We note that the prior models are required to result in meaningful solutions for the crust, that is, the minimisation procedure (see Sect.~\ref{sec:crust}) leads to positive neutron-gas and cluster densities.
The constraints considered in this work are the following:
\begin{enumerate}
    \item Nuclear masses. 
    For each set of parameters, $\mathbf{X_{\rm nuc}}$, the surface parameters in the CLD model are fitted to the AME 2020 \citep{ame2020} experimental masses. Parameter sets that yield a (non-)convergent fit are retained (discarded); thus, the corresponding pass-band filter $\omega_{\rm AME}(\mathbf{X})=\omega_{\rm AME}(\mathbf{X}_{\rm nuc})$ is set to 1 (0).
    The probability of each model is then quantified by the quality of the reproduction of the nuclear masses, 
    \begin{equation}
        P(c_{\rm AME}|\mathbf{X}) = \omega_{\rm AME} \ e^{-\chi^2(\mathbf{X})/2} \ ,
        \label{eq:filter-masses}
    \end{equation}
    with 
    \begin{equation}
        \chi^2(\mathbf{X}) = \frac{1}{N_{\rm dof}} \sum_{j=1}^{N_{\rm AME}} \frac{\left( M^{(j)}_{\rm theo,CLD}(\mathbf{X}) - M^{(j)}_{\rm AME} \right)^2}{\sigma_{\rm th}^2} \ ,
        \label{eq:chi2-masses}
    \end{equation}
    where the sum runs over the nuclei in the AME2020 mass table with $N,Z \ge 8$ ($N=A-Z$), $M_{\rm theo,CLD}(\mathbf{X})=M_{\rm theo,CLD}(\mathbf{X_{\rm nuc}})$ is the theoretical mass calculated with the CLD model using the parameter set $\mathbf{X}_{\rm nuc}$ complemented with the best-fit surface parameters, $M_{\rm AME}$ is the experimental mass, $\sigma_{\rm th} = 0.04$~MeV/$c^2$ is a systematical theoretical error, and $N_{\rm dof} = N_{\rm AME} - 4$ is the number of degrees of freedom. 
    This constraint is always applied in all distributions, including the prior.

    \item Chiral-EFT calculations.
    At low density (LD), models are selected using a strict filter from the chiral-EFT calculations. The energy per nucleon of symmetric nuclear matter and pure neutron matter predicted by each model are compared with the corresponding energy bands of \citet{Drischler2016} (see also Fig.~\ref{fig:comp-order-hm}), enlarged by $5\%$. This constraint is applied in the low-density region, from $0.02$~fm$^{-3}$ to $0.2$~fm$^{-3},$ or to the baryon density at which the piecewise polytropic EoS is matched if $n_{\rm match} < 0.2$~fm$^{-3}$. 
    The associated probability is given by
    \begin{equation}
        P(c_{\rm EFT}|\mathbf{X}_{\rm nuc}) = \omega_{\rm LD} \ ,
        \label{eq:filter-EFT}
    \end{equation}
    with $\omega_{\rm LD}(\mathbf{X}_{\rm nuc}) = 1\ (0)$ if the model given by the parameter set $\mathbf{X}_{\rm nuc}$ is (not) consistent with the EFT bands. 
    This constraint is always applied, except for the prior distributions.
    
    \item Causality and thermodynamic stability.
    This filter, which is effective at high density (HD), is similar to the low-density one, that is, only models satisfying causality and thermodynamical stability are allowed, and it reads
    \begin{equation}
        P(c_{\rm HD}|\mathbf{X}) = \omega_{\rm HD} \ ,
        \label{eq:filter-HD}
    \end{equation}
    with $\omega_{\rm HD}(\mathbf{X}) = 1\ (0)$ if the model is retained (discarded).
    Since the piecewise polytropic EoS is employed at high density, the thermodynamic stability, given by the condition $dP/d\rho_B \ge 0$, is guaranteed by construction for the polytropic segments (but could be violated before). As for causality, the polytropic EoS is calculated until the maximum density compatible with sub-luminal speed of sound.
    An additional check is performed on the meta-model part of the EoS; that is, the symmetry energy has to be non-negative at all densities where the meta-model is applied.
    This constraint is always applied, except for the prior distributions.

    \item NS maximum mass.
    This filter accounts for observational data of massive NSs, specifically of the precisely measured mass of PSR J0348$+$0432, $M_{J0348} = 2.01 \pm 0.04$\;M$_{\odot}$, with $\sigma_{J0348}=0.04$ being the standard deviation \citep{Antoniadis2013}. The associated likelihood is defined as the cumulative Gaussian distribution function,
    \begin{equation}
        P(c_{J0348}|\mathbf{X}) = \frac{1}{0.04 \sqrt{2 \pi}} \int_0^{M_{\rm max}(\mathbf{X})/M_\odot} e^{-\frac{(x-M_{J0348}/M_\odot)^2}{2 \sigma_{J0348}^2}} dx \ ,
        \label{eq:filter-Mmax}
    \end{equation}
     where $M_{\rm max}(\mathbf{X})$ is the maximum NS mass at equilibrium (compatible with causality), determined for each model from the solution of the TOV equations.
     This constraint is always applied, except for the prior distributions.
    
    \item LIGO-Virgo tidal deformability data from GW170817.
    The likelihood associated with the tidal deformability from the GW170817 event is written as
    \begin{equation}
        P(c_{\rm GW}|\mathbf{X}) = \sum_i P_{\rm GW}(\tilde{\Lambda}(q^{(i)}),q^{(i)}) \ ,
        \label{eq:filter-tidal}
    \end{equation}
    where $P_{\rm GW}(\tilde{\Lambda}(q^{(i)}),q^{(i)})$ is the joint posterior distribution of $\Lambda$ and $q$ \citet{Abbott2019prx}\footnote{Data of the posterior distributions are taken from the LIGO Document P1800061-v11 `Properties of the binary neutron star merger GW170817' at \url{https://dcc.ligo.org/LIGO-P1800061/public}, obtained using the PhenomPNRT waveform, referred to as `reference model' in \citet{Abbott2019prx}, and low-spin prior.}.
    In this work, $q$ is chosen to be in the one-sided $90\%$ confidence interval $q \in [0.73, 1.00]$. 
    Since the chirp mass in GW170817, $\mathcal{M}_{\rm c} = 1.186 \pm 0.001 M_\odot$, has been precisely determined for each value of the mass ratio, we calculate $m_1$ directly from the median value of $\mathcal{M}_{\rm c}$ using Eq.~\eqref{eq:mchirp} \citep{Abbott2019prx}.
    
    \item NICER data.
    The NICER likelihood probability is given by
    \begin{eqnarray}
        P(c_{\rm NICER}|\mathbf{X}) = \sum_i P_{\rm NICER-J0030}(M_{J0030}^{(i)},R_{J0030}^{(i)}) \nonumber \\
        \times \sum_j P_{\rm NICER-J0740}(M_{J0740}^{(j)},R_{J0740}^{(j)}) \ ,
        \label{eq:filter-nicer}
    \end{eqnarray}
    where $P_{\rm NICER-J0030}(M,R)$ ($P_{\rm NICER-J0740}(M,R)$) is the two-dimensional probability distribution of mass and radius for the pulsar PSR J0030$+$0451 (PSR J0740$+$6620) obtained using NICER (NICER and XMM-Newton) data and the waveform model with three uniform oval spots by \citet{Miller2019} (\citet{Miller2021}). The intervals of ${M_{J0030} = [1.0, 2.2]} M_\odot$ and ${M_{J0740} = [1.68,2.39]} M_\odot$ are chosen to be sufficiently large so that they cover the associated joint mass-radius distributions (see also Fig.~7 in \citet{Miller2019} and Fig.~1 in \citet{Miller2021}). More recently, \citet{Salmi2022} performed an analysis of PSR J0740$+$6620 from NICER data within a framework similar to that adopted in \citet{Miller2021}, leading to results ---in particular the inferred mass and radius--- consistent with previous findings and having similar uncertainties. Therefore, we still use the constraints from \citet{Miller2021}, as was done in \citet{Dinh2021c}.
    
\end{enumerate}

\begin{figure}
    \centering
    \includegraphics[width=\linewidth]{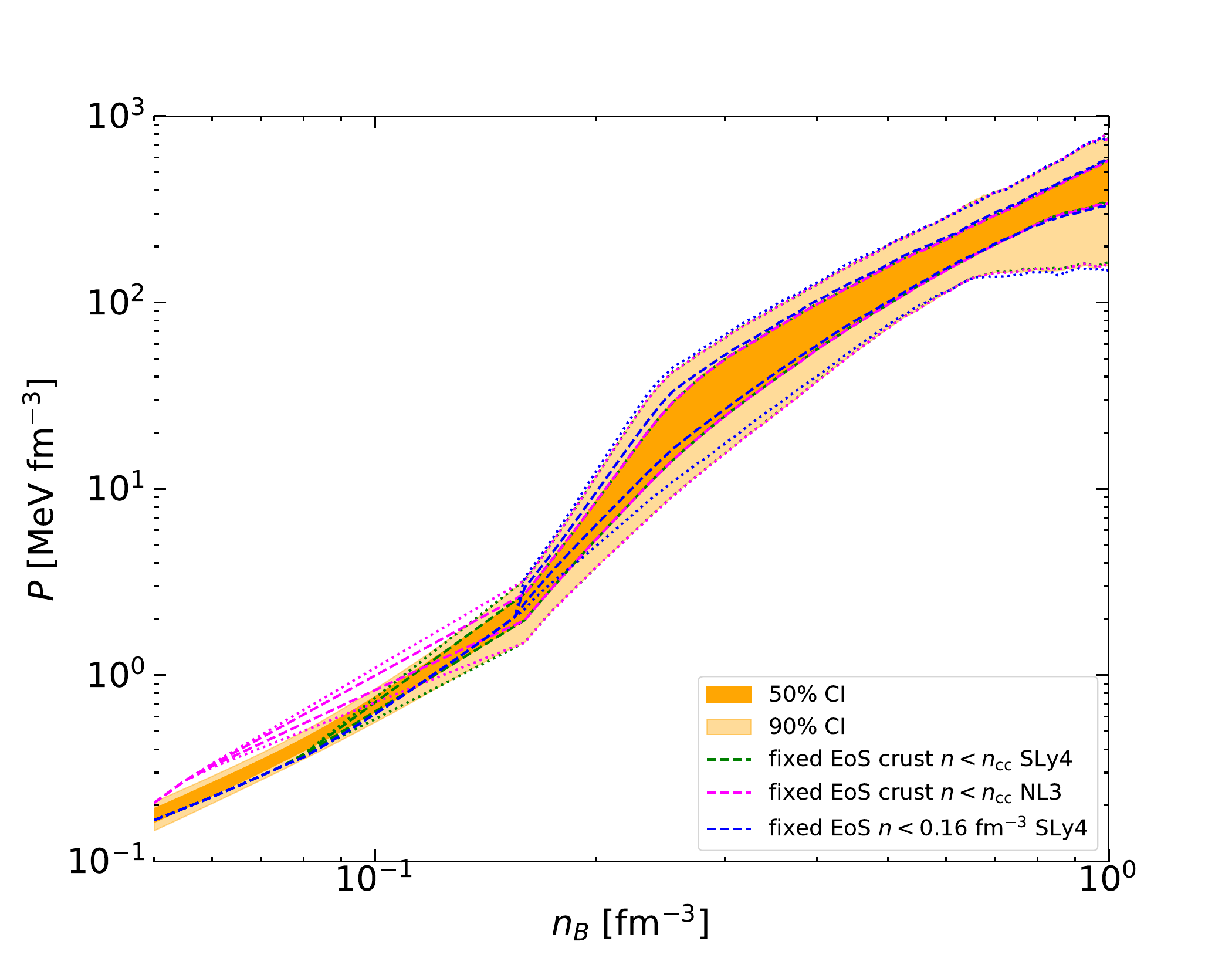}
    \caption{Posterior distribution for EoS. Dark and light yellow bands correspond to the $50\%$ and $90\%$ confidence intervals for the EoS, respectively, obtained including the low- and high-density (LD+HD) filters. Green (magenta) curves indicate the $50\%$ (dashed lines) and $90\%$ (dotted lines) confidence intervals when the crust based on SLy4 (NL3) is used, while blue lines correspond to using the EoS model based on SLy4 until ${n_{\rm match}=0.16}$~fm$^{-3}$ (see text for details).}
    \label{fig:eos}
\end{figure}

To perform the Bayesian analysis, $10^8$ ($1.5 \times 10^8$) parameter sets $\mathbf{X}$ are generated if $n_{\rm match} = 0.16$~fm$^{-3}$ ($n_{\rm match} = 0.32$~fm$^{-3}$), so as to have comparable statistics of the order of $10^4$ models for the posterior sets\footnote{The reason for generating a higher number of parameter sets if the polytropes are matched at $n_{\rm match} \ge 0.2$~fm$^{-3}$ reside in the stricter filter operated by the chiral-EFT. Indeed, the latter filter acts from $0.02$~fm$^{-3}$ to $0.16$~fm$^{-3}$ ($0.2$~fm$^{-3}$) if the polytrope is matched at $0.16$~fm$^{-3}$ ($ 0.32$~fm$^{-3}$).}.
In order to assess the impact of using a unique crust instead of a consistent EoS, for a given matching density $n_{\rm match}$, we run the following calculations:
\begin{enumerate}[(i)]
    \item we first construct a series of EoSs by matching to each crust (built with a parameter set $\mathbf{X}_{\rm nuc}$ complemented with the associated surface parameters; see Sect.~\ref{sec:crust}) plus part of the core (built with the meta-model approach using the parameter set $\mathbf{X}_{\rm nuc}$) a piecewise-polytrope EoS; to each model, the low- and high-density (LD+HD) filters, including the maximum mass filter, Eqs.~\eqref{eq:filter-masses}, \eqref{eq:filter-EFT}, \eqref{eq:filter-HD}, and \eqref{eq:filter-Mmax}, are applied;
    \item using the resulting high-density part of the EoS, that is, the resulting piecewise polytrope parameter sets $\mathbf{X}_{\rm poly}$ from point (i), we glue a unique low-density EoS. Two scenarios are considered. These are as follows.
    \begin{enumerate}[(a)]
        \item A unique crust EoS is calculated until the corresponding crust--core transition density, $n_{\rm cc}$. Here, an extra polytrope is added, with the requirement of the pressure continuity at the transition between $n_{\rm cc}$ and $n_{\rm match}$. At this point, the piecewise polytrope resulting from the calculations (i) is matched. This construction ensures a continuous and smooth EoS.
        \item A unique EoS (crust plus part of the core) is calculated until $n_{\rm match}$, at which point the piecewise polytrope resulting from the calculations in point (i) is matched. In this case, we ensure the monotonicity of pressure at $n_{\rm match}$, but continuity is not necessarily guaranteed.
    \end{enumerate}
\end{enumerate}
To illustrate this procedure, we show in Fig.~\ref{fig:eos} an example of the EoS models obtained for the value of ${n_{\rm match}=0.16}$~fm$^{-3}$: the yellow shaded areas represent the $50\%$ (dark) and $90\%$ (light) confidence intervals for the EoS calculated as described in point (i), while the curves correspond to the $50\%$ (dashed lines) and $90\%$ (dotted lines) confidence intervals for the EoS models built as per point (ii).
 We chose as unique EoSs (crust plus eventually part of the core) those based on the SLy4 functional, for which the green (blue) curves correspond to the EoS constructed as described in point ii.a (point ii.b), and on the NL3 (magenta curves) functional as illustrative examples. 
The unique EoS models based on the SLy4 and NL3 functionals were obtained within the CLD model, as described in Sect.~\ref{sec:crust}. 
The empirical parameters entering Eq.~\eqref{eq:meta} for these functionals are given in Table~1 of \citet{Dinh2021a}; with these parameter values, the meta-model approach indeed gives a very good reproduction of the SLy4 and NL3 functionals.
We note that for NL3 it is not possible to build an EoS until $n_{\rm match}=0.16$~fm$^{-3}$; indeed, this EoS model is too stiff to be compatible with our posterior (yellow band) thus no polytropic EoS can be matched monotonically at this density. 
We nevertheless show this EoS as we consider it an extreme case.
The choice of SLy4 was instead made because a unique low-density EoS from this functional following the work by \citet{Douchin01} is often employed in Bayesian inference analyses (see e.g. \citet{Essick2020prc}).
As one can see, at high density the EoSs coincide, by construction.
When the procedure (ii.a) is used, the additional polytrope built between the crust--core transition and $n_{\rm match}$ allows the EoS to remain smooth and to retain part of the outer-core uncertainties in the EoS.
On the other hand, when the unique EoS is glued as described in (ii.b), jumps in pressure are possible at $n_{\rm match}$, although these remain relatively small because our initial posterior (yellow bands) is already relatively narrow (mainly due to the low-density EFT filter) and the EoS based on SLy4 agrees well with it.

\begin{figure*}
    \centering
    \includegraphics[scale=0.3]{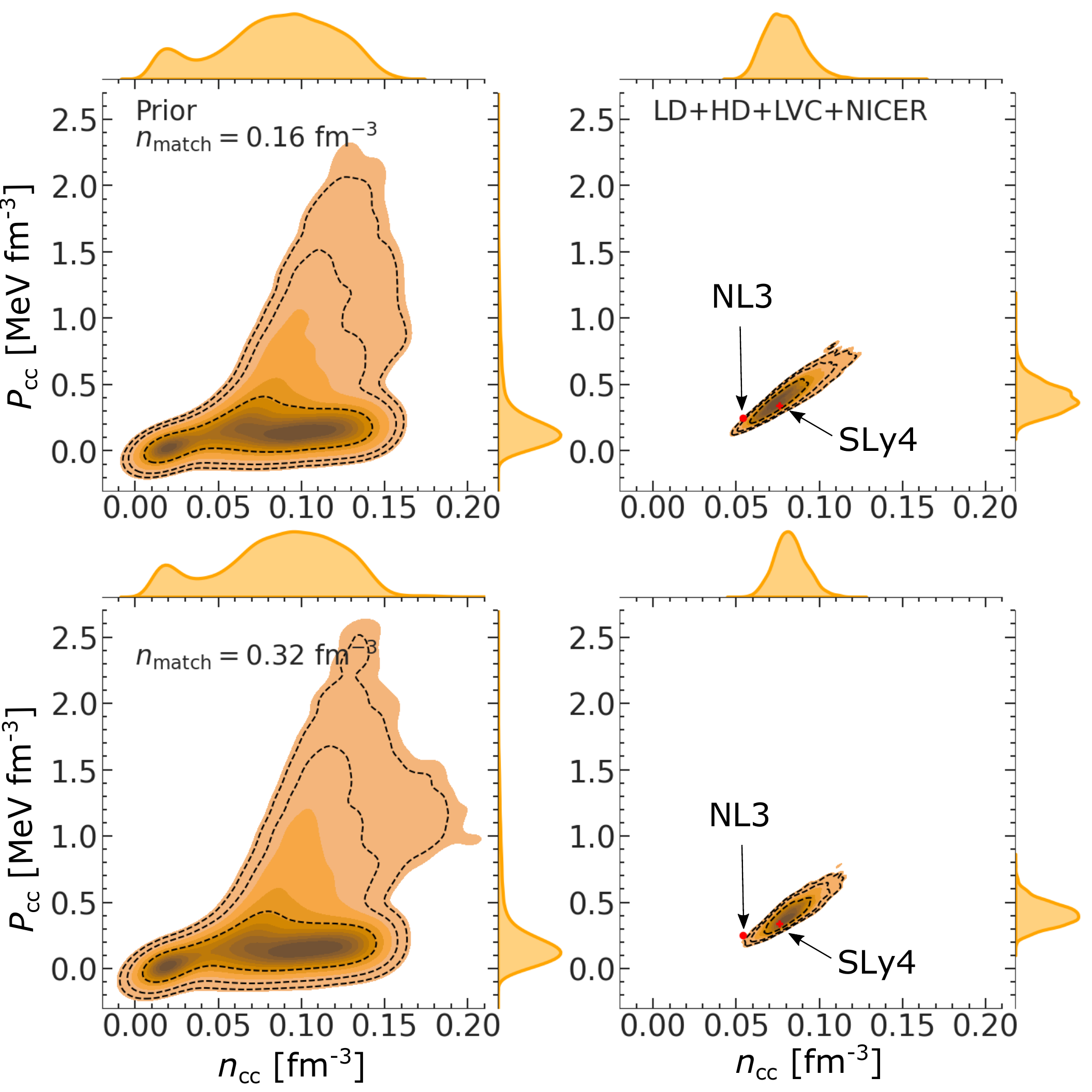}
    \caption{Prior (left panels) and posterior (right panels) distributions for crust--core transition when the piecewise polytrope EoS is matched at $n_{\rm match} = 0.16$~fm$^{-3}$ (top panels) and at $n_{\rm match} = 0.32$~fm$^{-3}$ (bottom panels). Dashed black curves represent the $1\sigma$, $2\sigma$, and $3\sigma$ confidence levels. Low- and high-density filters together with constraints from GW170817 and NICER (LD+HD+LVC+NICER) are applied. The red cross (circle) indicated by arrows shows the crust--core transition predicted by the EoS based on SLy4 (NL3). 
    }
    \label{fig:cc}
\end{figure*}

A complementary way to analyse the impact of the use of a unique crust on the EoS prediction is by looking at the crust--core transition.
In Fig.~\ref{fig:cc}, we show the joint distribution of the density and pressure at the crust--core transition, obtained when the polytropes are matched at either $0.16$~fm$^{-3}$ (top panels) or at $0.32$~fm$^{-3}$ (bottom panels).
The prior distributions are shown in the left panels, while the posterior distributions obtained applying the LD+HD filters together with the filters from GW170817 and NICER (LD+HD+LVC+NICER; see Eqs.~\eqref{eq:chi2-masses}, \eqref{eq:filter-EFT}, \eqref{eq:filter-HD}, \eqref{eq:filter-Mmax}, \eqref{eq:filter-tidal}, and \eqref{eq:filter-nicer}) are shown in the right panels.
The combined filters considerably reduce the spread in the predictions of the crust--core transition, which is thus relatively well determined.
However, the choice of a unique crust completely erases the uncertainties since only a unique value of the crust--core transition point is defined for a given model.
This can be seen by comparing the shaded yellow region in the right panels of Fig.~\ref{fig:cc}, with the two symbols representing the crust--core transition for the EoS based on the SLy4 model (red plus symbol) and the NL3 model (red circle) as illustrative examples.
Incidentally, the NL3 model is only very marginally compatible with our posterior if the polytrope is matched at $0.16$~fm$^{-3}$, as already noticed for the EoS (see Fig.~\ref{fig:eos}), while it is incompatible if the polytrope is matched at $0.32$~fm$^{-3}$.
This is because in the latter case the stricter EFT filter applied until $0.2$~fm$^{-3}$ excludes too stiff models such as NL3.

\begin{figure}
    \centering
    \includegraphics[width=.95\linewidth]{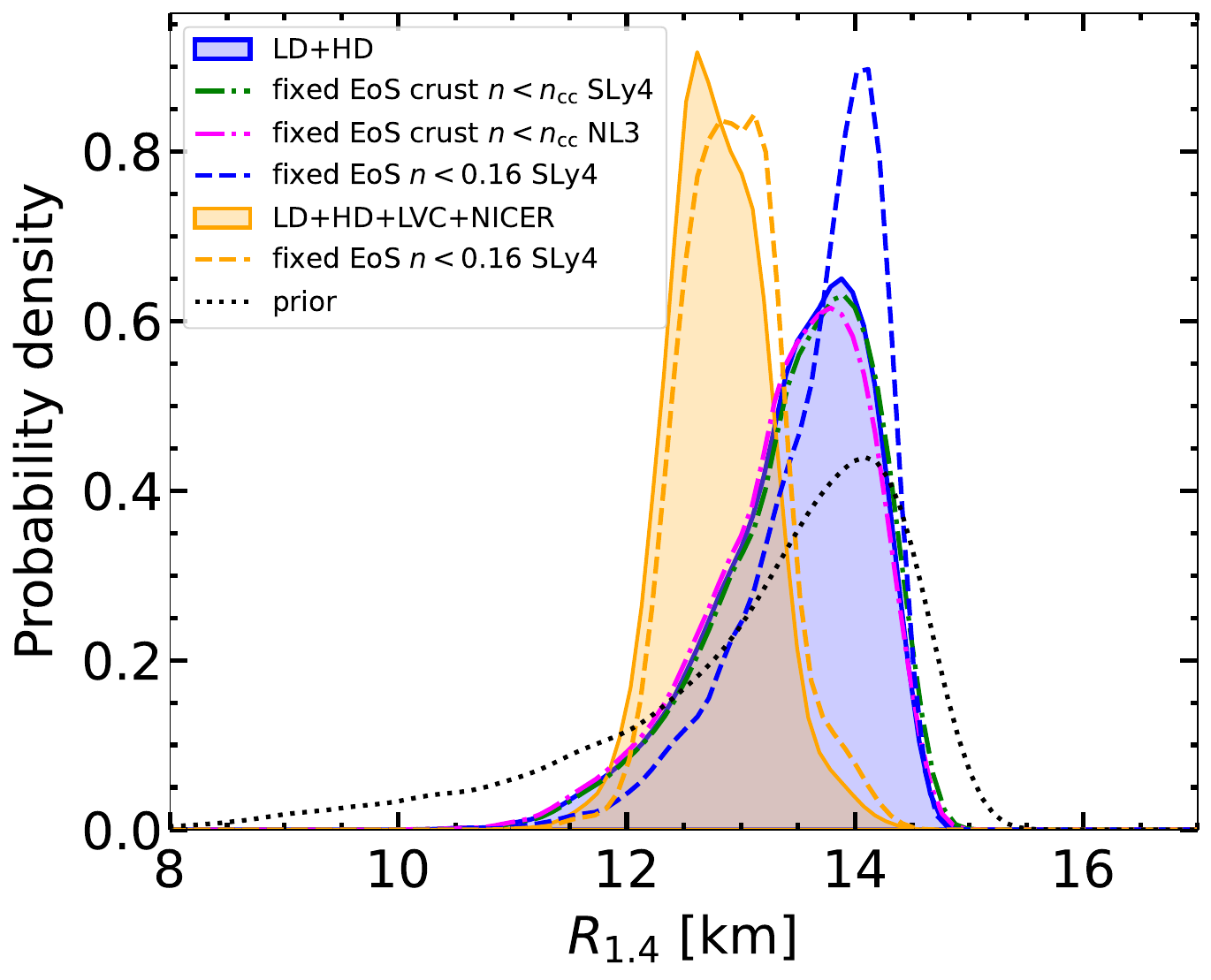}
    \caption{Posterior distribution for radius of a $1.4 M_\odot$ NS. The blue (orange) shaded area corresponds to the posterior obtained by applying the low- and high-density (LD+HD) filters (combined with the constraints from GW170817 and NICER). The blue and orange dashed curves are obtained by employing a unique EoS based on SLy4 until $0.16$~fm$^{-3}$, while the dash-dotted curves are obtained by employing the crust EoS based on SLy4 (green) and NL3 (magenta) until the crust--core transition. For comparison, the prior is shown by a dotted black line (see text for details).}
    \label{fig:R14_016}
\end{figure}

We now discuss the impact of the use of a unique instead of a consistent EoS on more global NS observables.
We first show in Fig.~\ref{fig:R14_016} the posterior distribution for the radius of a $1.4 M_\odot$ NS, $R_{1.4}$, as obtained using consistent EoSs (i.e. built as explained in point i), where the matching to a piecewise polytrope is done at $n_{\rm match} = 0.16$~fm$^{-3}$; the LD+HD filters are applied (blue shaded area).
The use of a unique crust below the crust--core transition (procedure described in point ii.a) does not change the average values of $R_{1.4}$ considerably, as can be seen by comparing the blue shaded area with the dash-dotted green (magenta) curves corresponding to the use of the crust EoS based on the SLy4 (NL3) functional.
However, when a unique model (in this case, SLy4, dashed blue curve) is applied until $n_{\rm match}$ (procedure described in point ii.b), although the average value is still not significantly affected, the width of the distribution is reduced, meaning that the use of a unique EoS underestimates the uncertainties.

Additionally, when the filters from GW170817 and NICER are included (orange shaded area), the distribution is shifted to lower radii, as expected since these measurements (and particularly the tidal deformability constraint from GW170817) tend to favour softer EoSs.
In this case, the effect of employing a unique EoS until $n_{\rm match}$ on the average and width of the distribution is reduced, as can be seen by comparing the shaded orange area with the dashed orange curve.
This can be understood from the fact that the constraints on the tidal deformability and the radius inferred from GW170817 and NICER observations favour specific values of the radii, thus shifting both distributions towards the same lower radii.
For comparison, the prior distribution is shown by a dotted black line; as expected, the prior distribution is much larger, and the applied filters tend to exclude too low (high) values of $R_{1.4}$ corresponding to softer (stiffer) EoSs.
This was already noticed in \citet{Dinh2021c}, where it was pointed out that lower values are filtered by the maximum mass constraint, which would exclude EoSs that are too soft, while higher values corresponding to stiff EoSs are essentially disfavoured by the EFT filter.

\begin{figure*}
\centering
\begin{minipage}[t]{.45\textwidth}
  \centering
  \includegraphics[width=.95\linewidth]{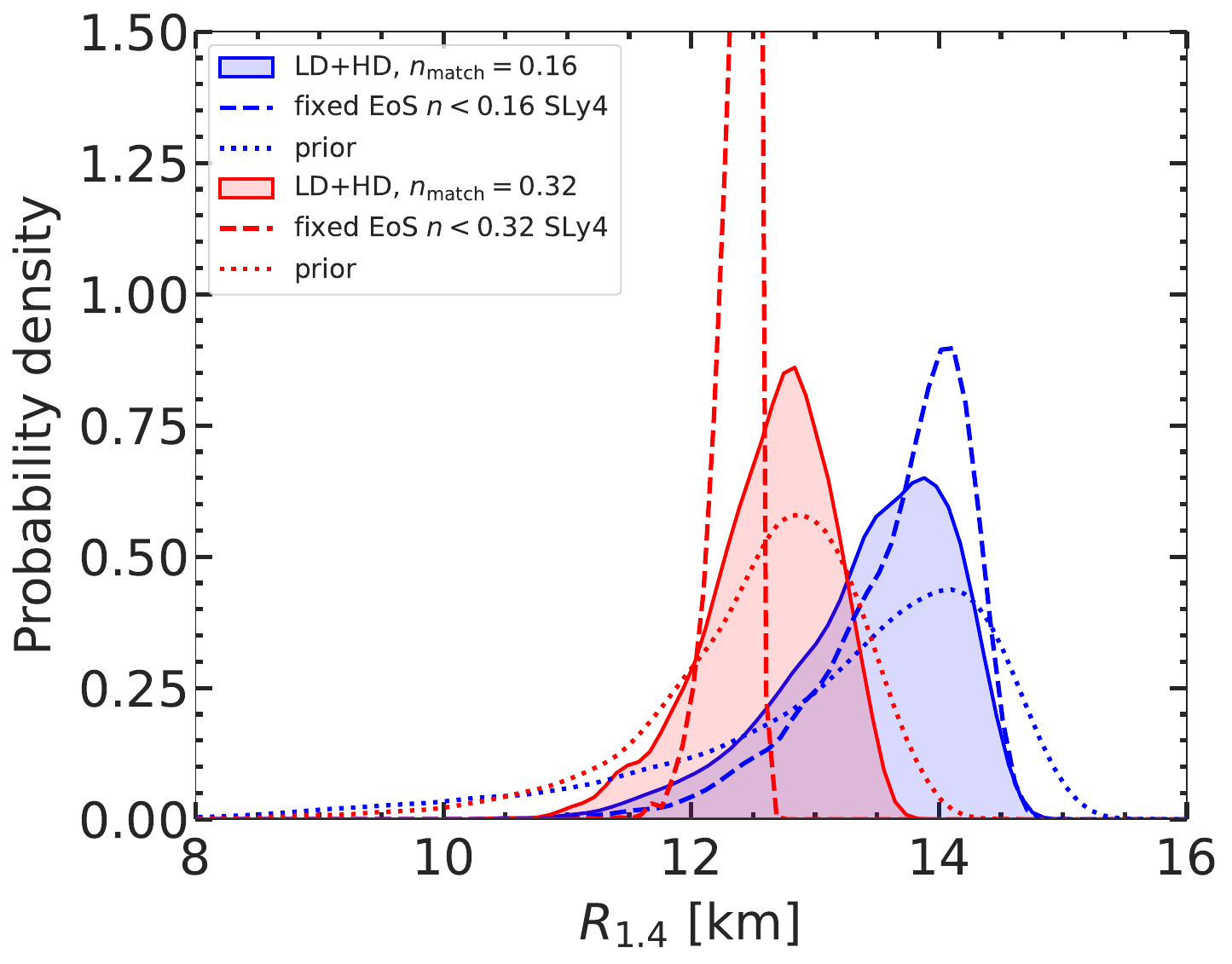}
\end{minipage} 
\hspace{0.1pc}
\begin{minipage}[t]{.45\textwidth}
  \centering
  \includegraphics[width=.95\linewidth]{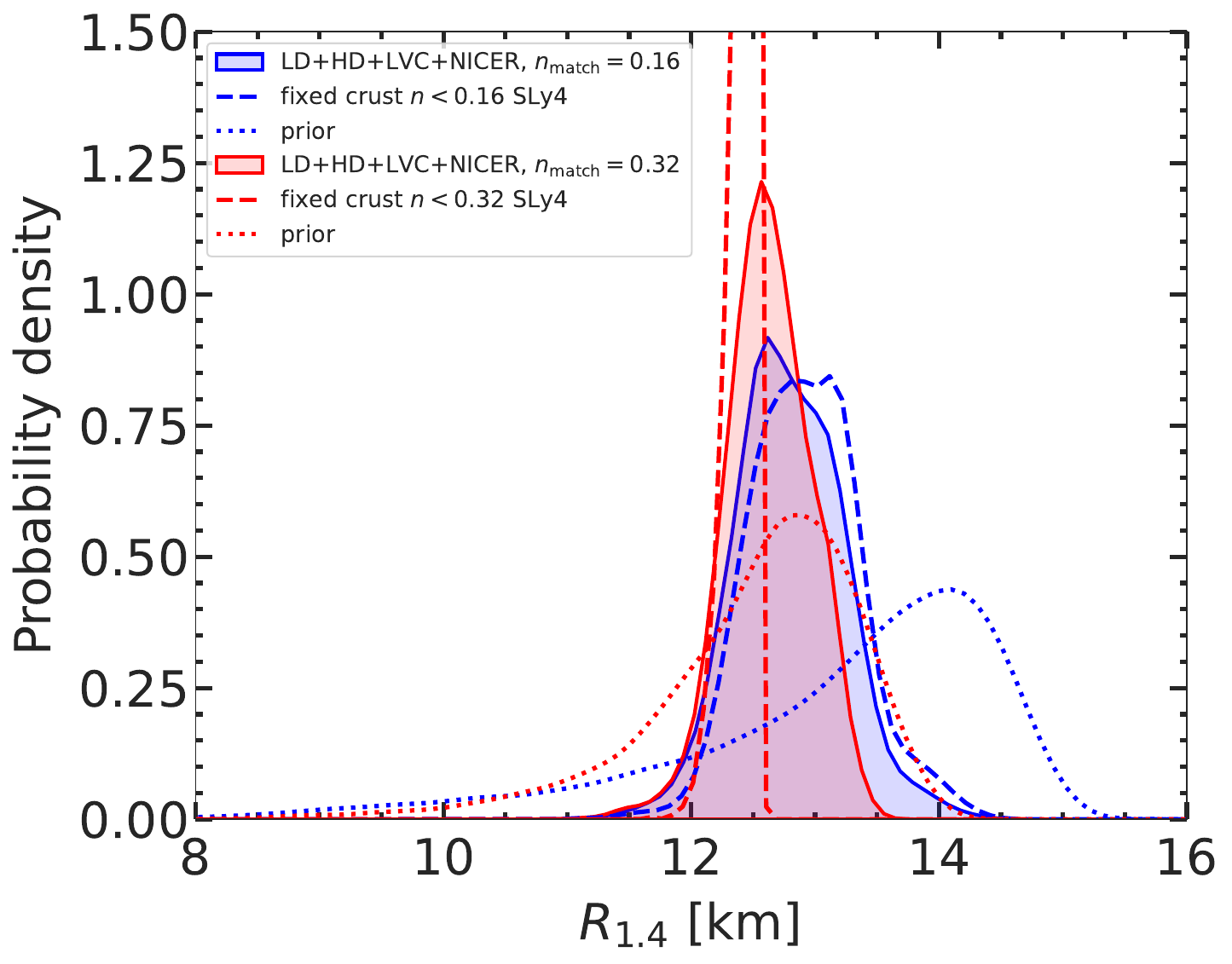}
\end{minipage}
\caption{Posterior distribution for radius of a $1.4 M_\odot$ NS. Left panel: Blue (red) shaded area corresponds to posterior obtained by applying the low- and high-density (LD+HD) filters and matching the polytropes at $0.16$~fm$^{-3}$ ($0.32$~fm$^{-3}$). The blue and red dashed curves are obtained by employing a unique EoS based on SLy4 until $0.16$~fm$^{-3}$ and $0.32$~fm$^{-3}$, respectively. For comparison, the corresponding priors are shown by dotted lines. 
Right panel: Same as left panel, but the posterior is obtained by applying the low- and high-density filters combined with the constraints from GW170817 and NICER (LD+HD+LVC+NICER;
see text for details).}
\label{fig:R14_LDHD}
\end{figure*}

\begin{figure*}
\centering
\begin{minipage}[t]{.45\textwidth}
  \centering
  \includegraphics[width=.95\linewidth]{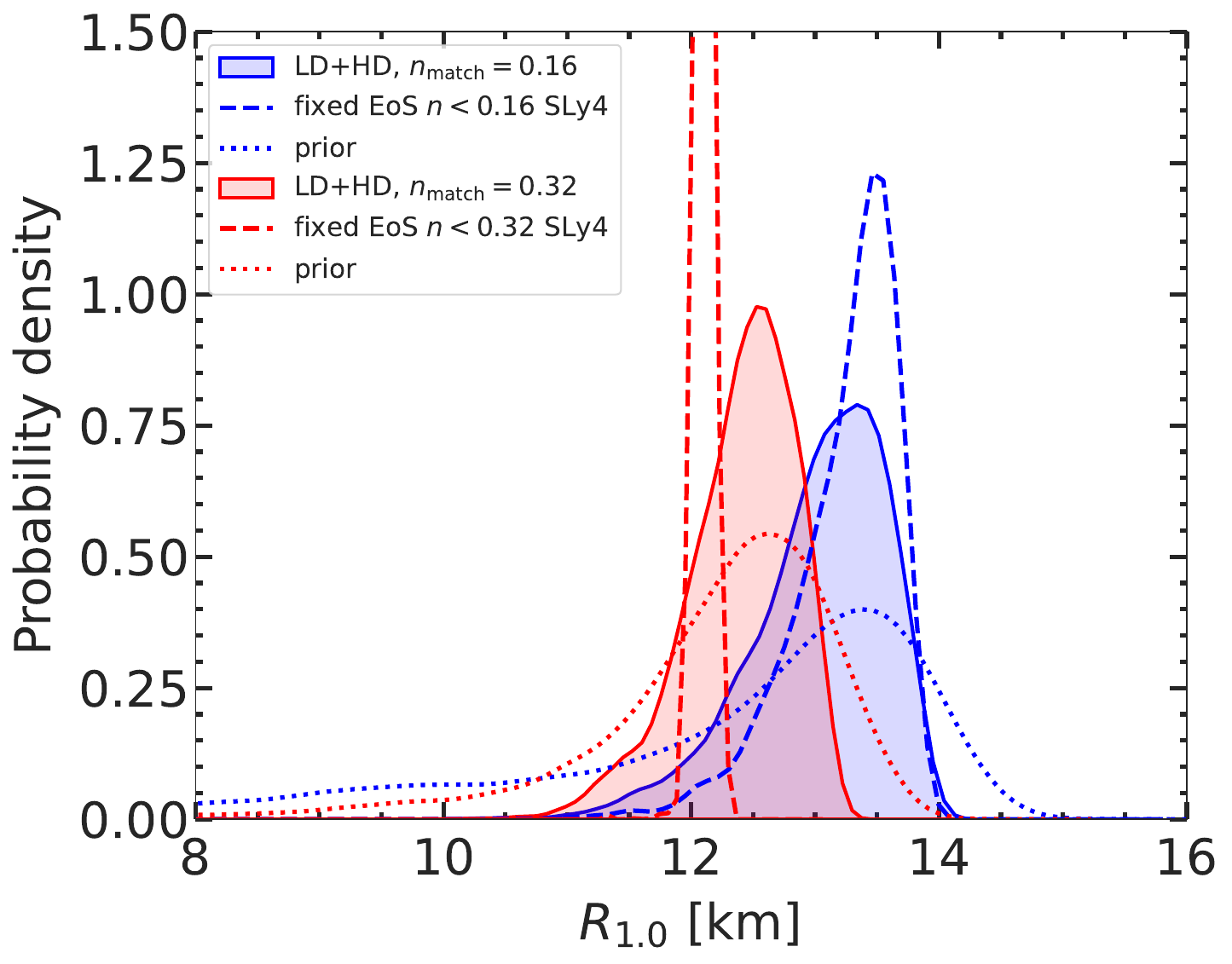}
\end{minipage} 
\hspace{0.1pc}
\begin{minipage}[t]{.45\textwidth}
  \centering
  \includegraphics[width=.95\linewidth]{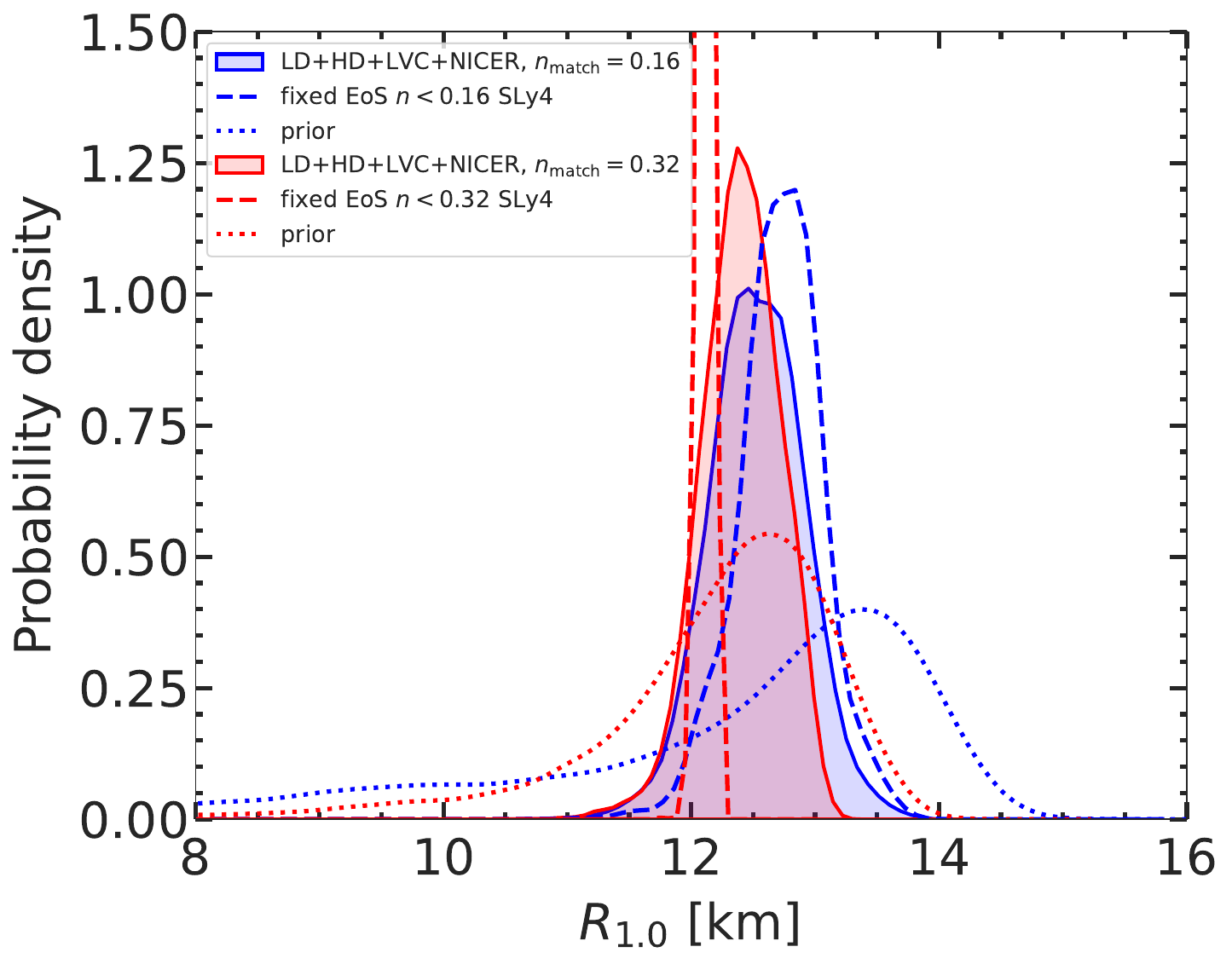}
\end{minipage}
\caption{Same as in Fig.~\ref{fig:R14_LDHD}, but for radius of a $1.0 M_\odot$ NS.}
\label{fig:R10_LDHD}
\end{figure*}

\begin{figure*}
\centering
\begin{minipage}[t]{.45\textwidth}
  \centering
  \includegraphics[width=.95\linewidth]{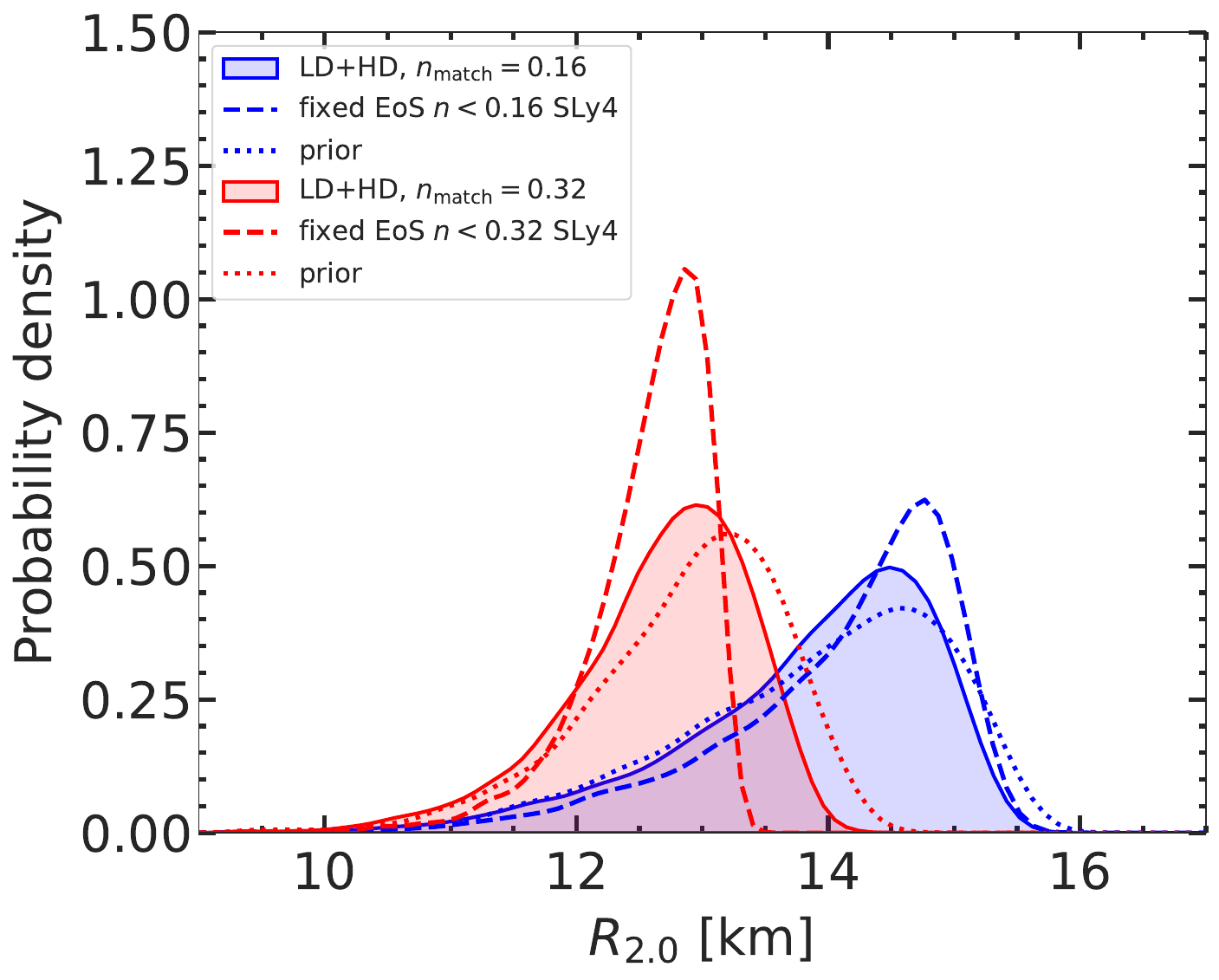}
\end{minipage} 
\hspace{0.1pc}
\begin{minipage}[t]{.45\textwidth}
  \centering
  \includegraphics[width=.95\linewidth]{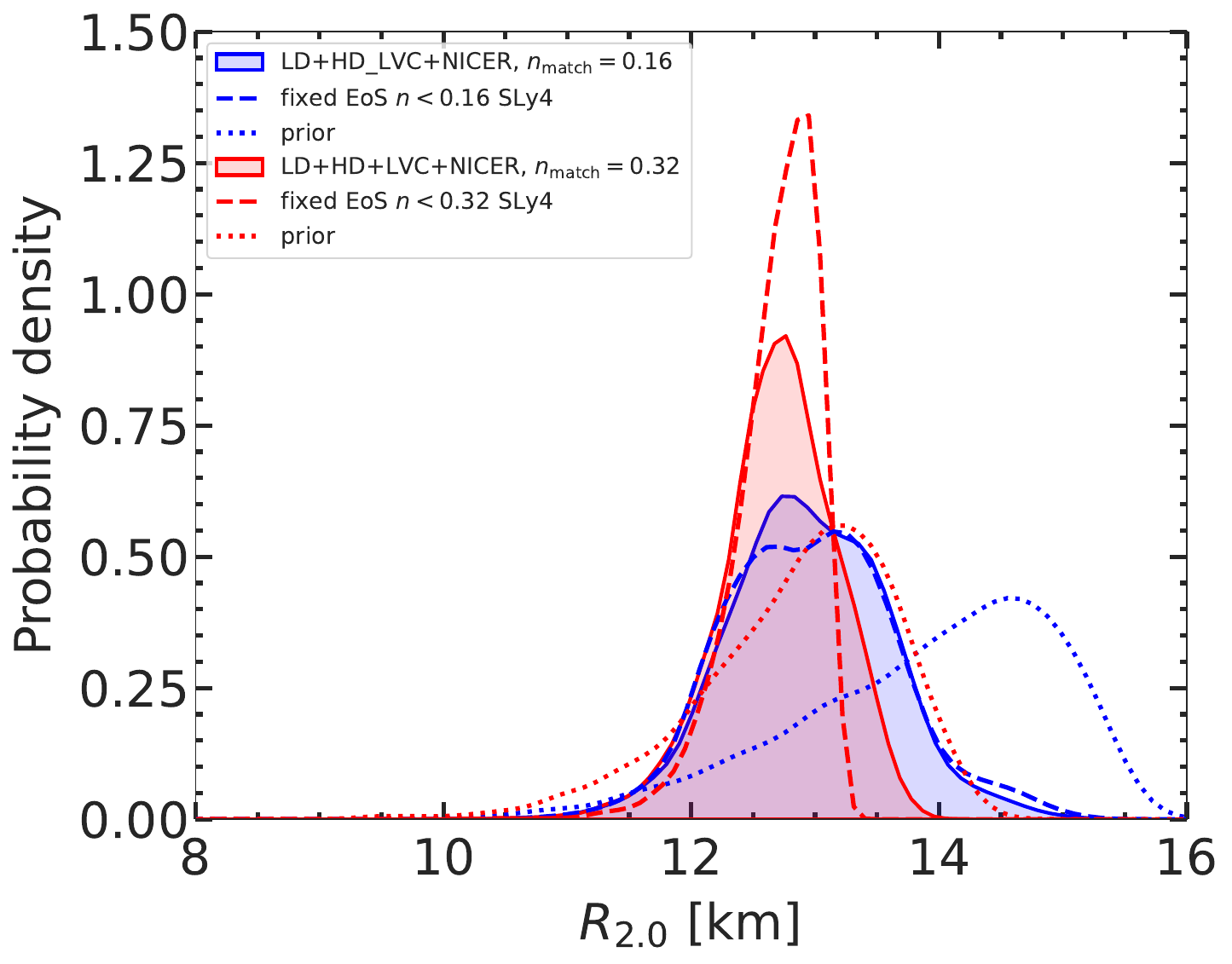}
\end{minipage}
\caption{Same as in Fig.~\ref{fig:R14_LDHD}, but for radius of a $2.0 M_\odot$ NS.}
\label{fig:R20_LDHD}
\end{figure*}

To explore the impact of different matching densities, in Fig.~\ref{fig:R14_LDHD} we show the comparison among the results on $R_{1.4}$ obtained with $n_{\rm match} = 0.16$~fm$^{-3}$ (blue curves) and $n_{\rm match} = 0.32$~fm$^{-3}$ (red curves).
The posterior distributions are obtained including the LD+HD filters (left panel) and the LD+HD filters together with the constraints from GW170817 on the tidal deformability and from NICER (right panel).
From both panels in Fig.~\ref{fig:R14_LDHD}, we can observe that the use of a unique model until $n_{\rm match}$ underestimates the uncertainties on the radius. 
This can be understood from the fact that in the latter case the unique (SLy4) model is applied until higher densities, thus constraining the predictions of the observables to those obtained with the chosen EoS.
Moreover, matching the polytropes at higher densities noticeably shifts the average value of $R_{1.4}$ to lower values and reduces the width of the distribution, even in the priors.
This can be due to two effects: first, if $n_{\rm match} = 0.32$~fm$^{-3}$, then no phase transition, mimicked by the sudden change of slope that can occur for the polytropic EoS, is allowed (and this is true even for the priors); and, secondly, the EFT filter, which tends to eliminate EoSs that are too stiff and predict higher values of the radius, is applied in a wider density region, thus further constraining the resulting distribution. 
Incidentally, a similar behaviour can be seen in Fig.~2 of \citet{Tews2018} and Fig.~3 of \citet{Tews2019}, although in the latter works a sound-speed model is used instead of a piecewise polytrope.
However, when the constraints from the LVC and NICER are applied, the shift in the predicted radius and the difference between the use of a unique EoS instead of a consistent one are decreased because, as already noticed in Fig.~\ref{fig:R14_016}, these filters tend to favour softer EoSs and thus constrain the radii.
The same trend is also observed for the radius of a $1.0 M_\odot$ and $2.0 M_\odot$ NS, as can be seen from Figs.~\ref{fig:R10_LDHD} and \ref{fig:R20_LDHD}, respectively.

\begin{figure*}
\centering
\begin{minipage}[t]{.45\textwidth}
  \centering
  \includegraphics[width=.95\linewidth]{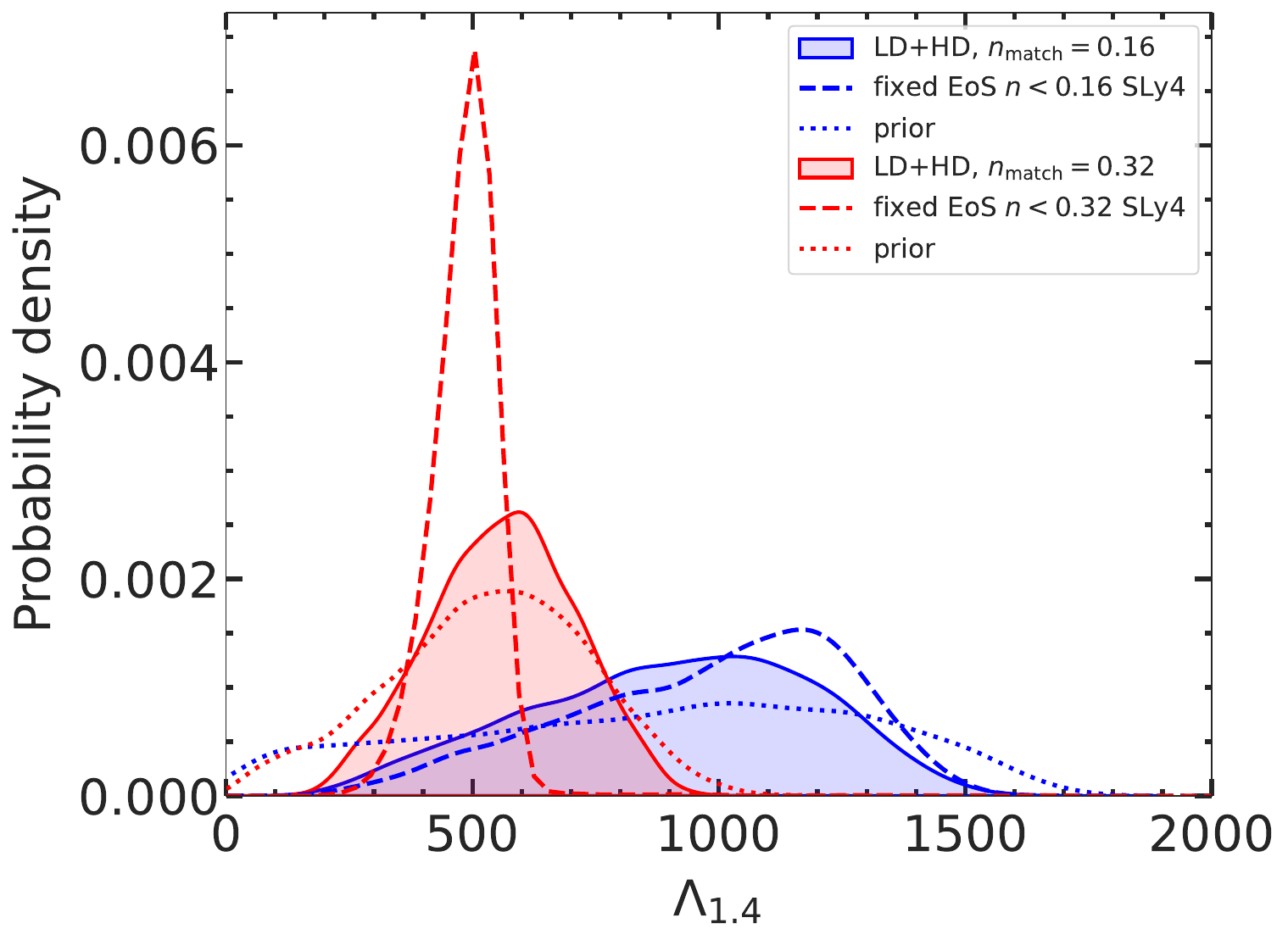}
\end{minipage} 
\hspace{0.1pc}
\begin{minipage}[t]{.45\textwidth}
  \centering
  \includegraphics[width=.95\linewidth]{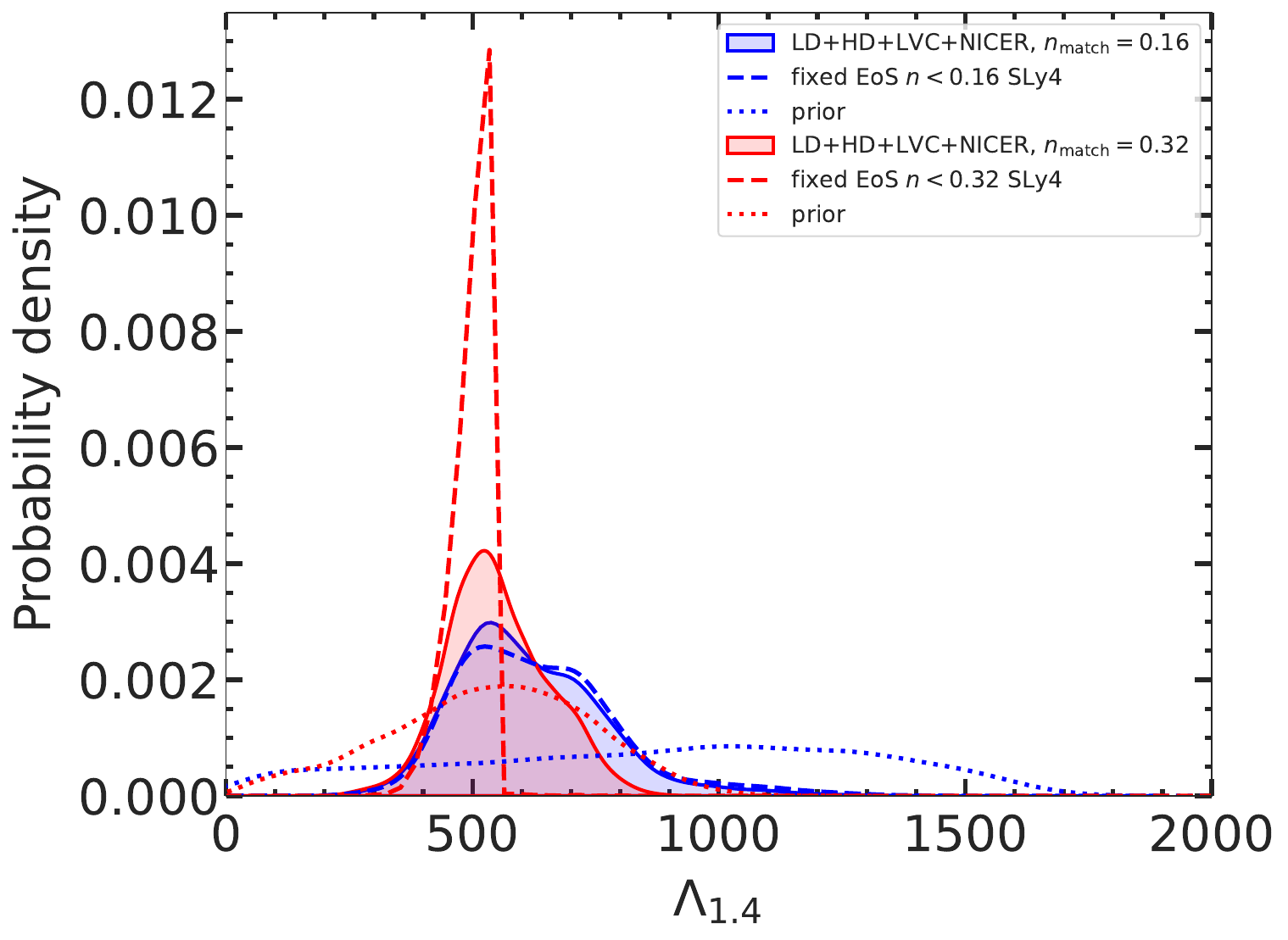}
\end{minipage}
\caption{Same as in Fig.~\ref{fig:R14_LDHD}, but for tidal deformability of a $1.4 M_\odot$ NS.}
\label{fig:Lambda14_LDHD}
\end{figure*}

\begin{figure*}
\centering
\begin{minipage}[t]{.45\textwidth}
  \centering
  \includegraphics[width=.95\linewidth]{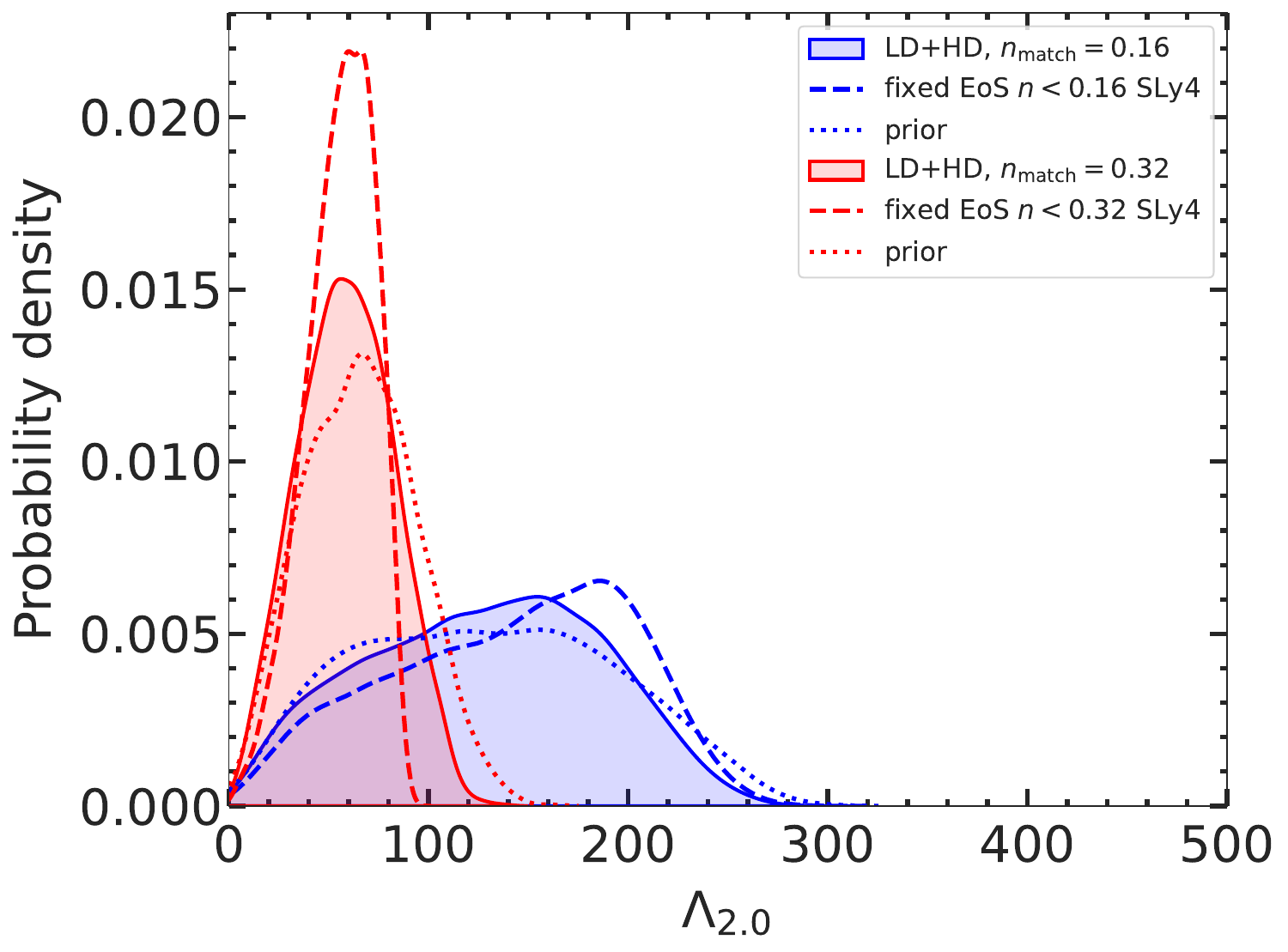}
\end{minipage} 
\hspace{0.1pc}
\begin{minipage}[t]{.45\textwidth}
  \centering
  \includegraphics[width=.95\linewidth]{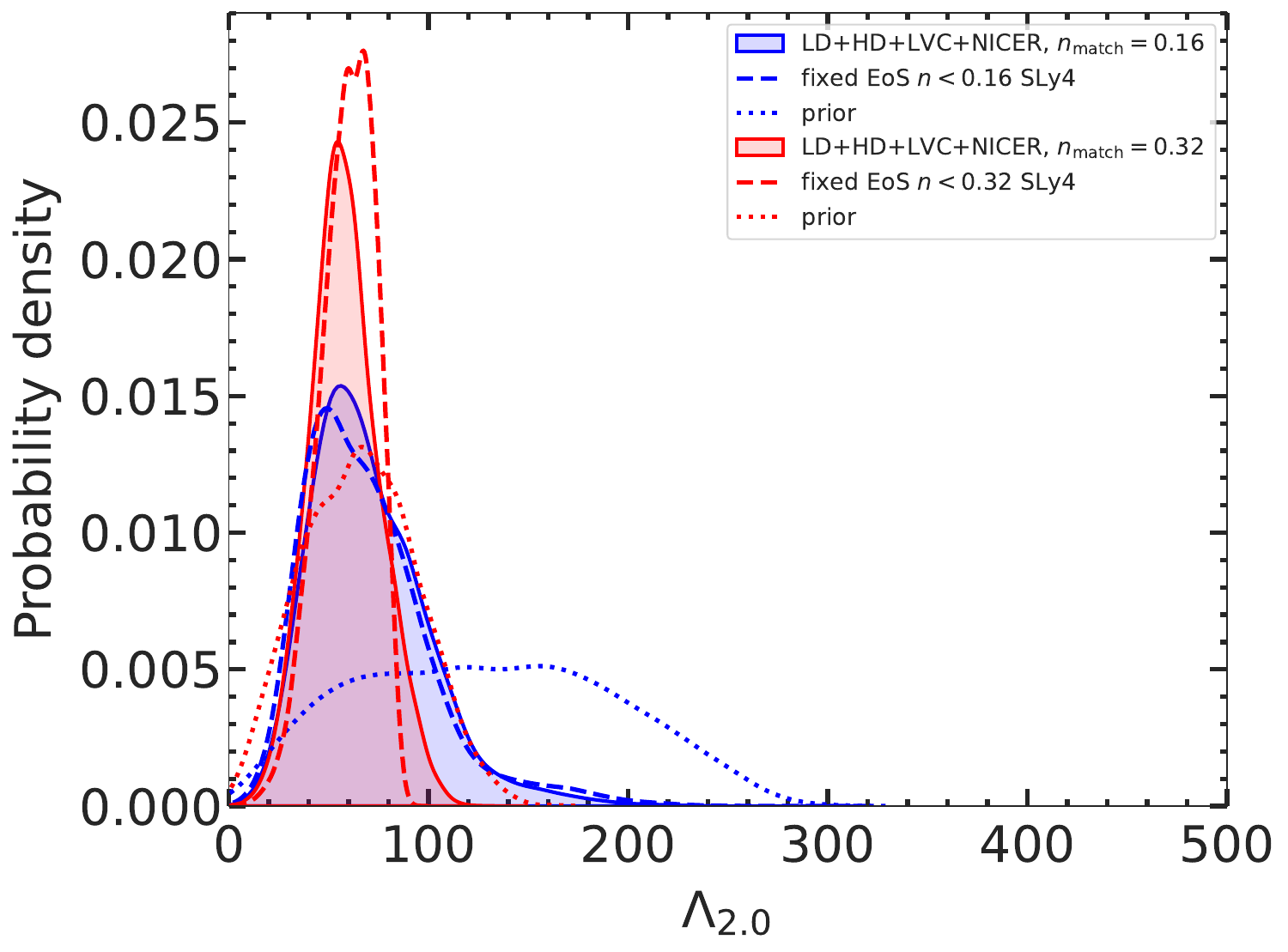}
\end{minipage}
\caption{Same as in Fig.~\ref{fig:R14_LDHD}, but for tidal deformability of a $2.0 M_\odot$ NS.}
\label{fig:Lambda20_LDHD}
\end{figure*}

A similar discussion applies for the tidal deformability of a $1.4 M_\odot$ and $2.0 M_\odot$ NS, illustrated in Figs.~\ref{fig:Lambda14_LDHD} and \ref{fig:Lambda20_LDHD}, respectively.
Here again, employing a unique EoS model has the effect of slightly shifting the average values and reducing the uncertainties (see left panels of Figs.~\ref{fig:Lambda14_LDHD} and \ref{fig:Lambda20_LDHD}), and the impact decreases when the filters on the tidal deformability inferred from GW170817 and NICER are applied (see right panels in Figs.~\ref{fig:Lambda14_LDHD} and \ref{fig:Lambda20_LDHD}).
An additional remark is that changing the matching density from $n_{\rm match} = 0.16$~fm$^{-3}$ to $n_{\rm match} = 0.32$~fm$^{-3}$ considerably decreases the width of the $\Lambda$ distribution, thus reducing the associated uncertainties.
Incidentally, a similar effect can be deduced from Fig.~1 in \citet{Tews2018}.

\begin{figure*}
\centering
\begin{minipage}[t]{.45\textwidth}
  \centering
  \includegraphics[width=.95\linewidth]{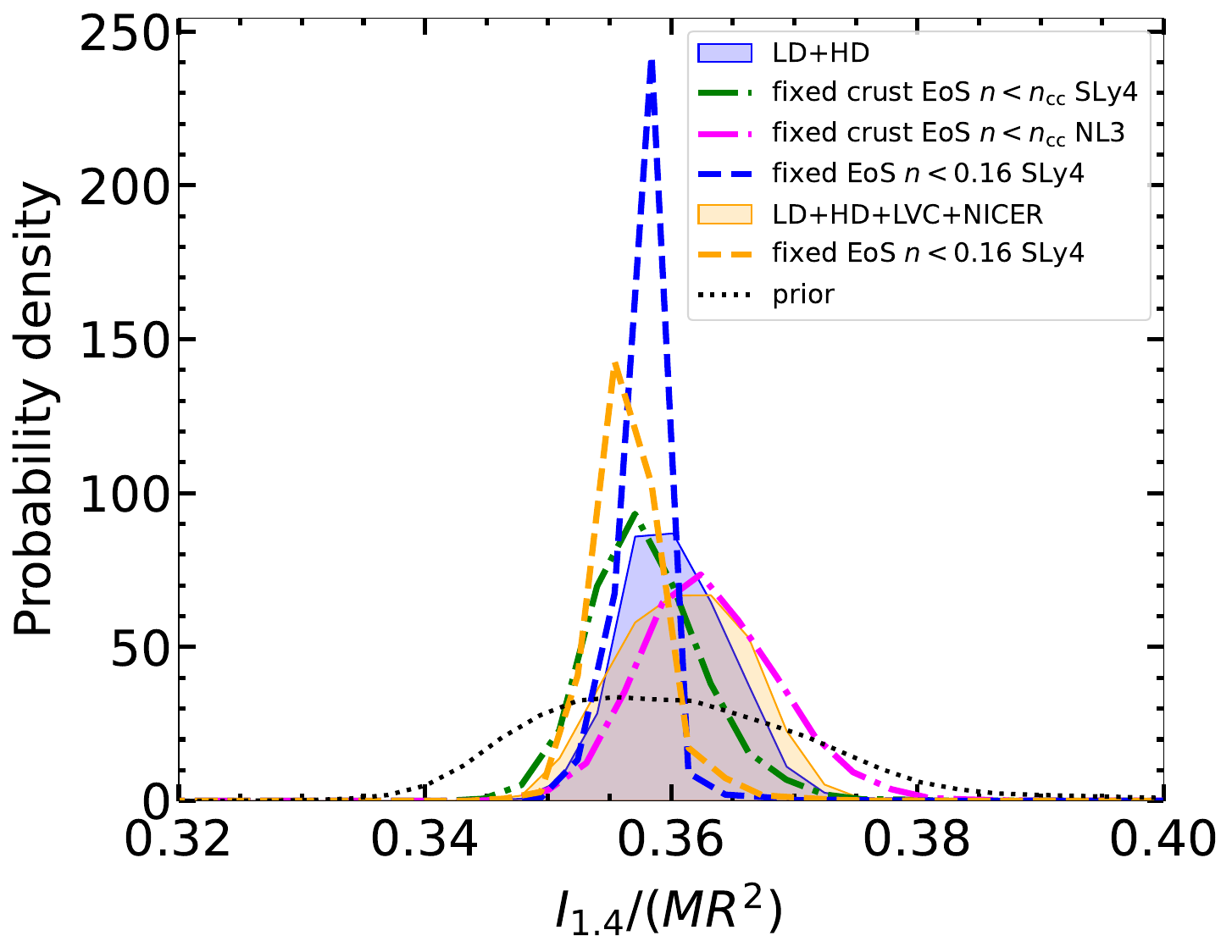}
\end{minipage} 
\hspace{0.1pc}
\begin{minipage}[t]{.45\textwidth}
  \centering
  \includegraphics[width=.95\linewidth]{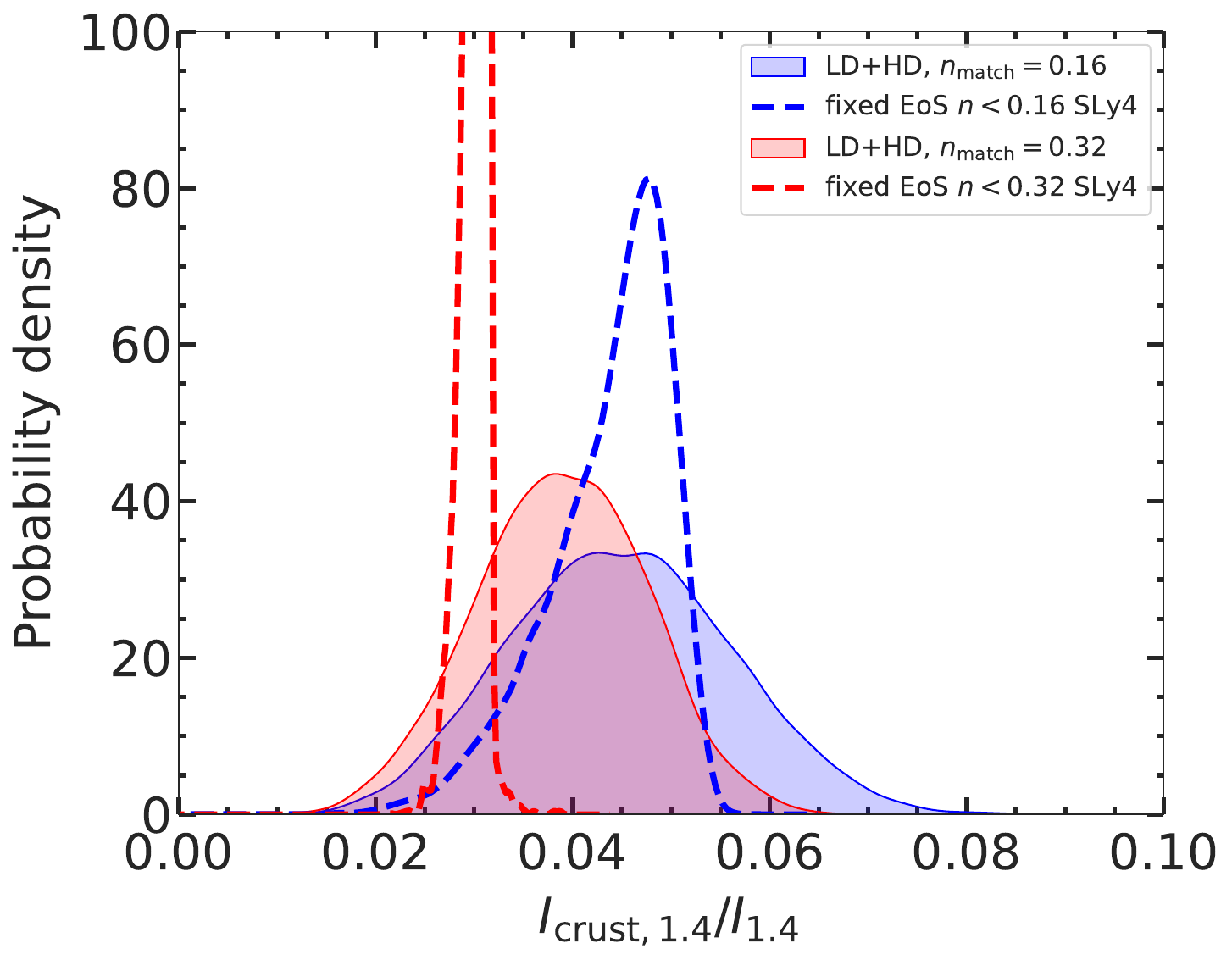}
\end{minipage}
\caption{Posterior distributions for moment of inertia of a $1.4 M_\odot$ NS. 
Left panel: Same as in Fig.~\ref{fig:R14_016}, but for normalised moment of inertia $I_{1.4}/(MR^2)$. 
Right panel: Same as in Fig.~\ref{fig:R14_LDHD}, but for moment of inertia of crust, $I_{{\rm crust}, 1.4}/I_{1.4}$ for $n_{\rm match} = 0.16$~fm$^{-3}$ (blue) and $n_{\rm match} = 0.32$~fm$^{-3}$ (red), when the LD+HD constraints are applied.}
  \label{fig:I14_LDHD}
\end{figure*}

Of particular interest for the NS modelling, and specifically for the glitch modelling, is the NS moment of inertia.
Predictions for the moment of inertia, as well as for the crustal thickness, have also been made, for example in \citet{Steiner2015}, although in the latter work a simple empirical expression was used to determine the crust--core transition, unlike in this work where it is consistently calculated for each given EoS.
In this work, the moment of inertia was calculated following \citet{Lattimer2016} (see their Eqs.~(84)-(85) and \citet{Hartle1967} for the seminal paper on the calculation of the moment of inertia in the slow rotation approximation; see also the discussion in \citet{Lattimer2000, Carreau2019b}): 
\begin{equation}
    I = \frac{c^2}{G} \frac{w(R) R^3}{6 + 2 w(R)} \ ,
    \label{eq:I}
\end{equation}
where $w(R)$ is the solution at the star radius $R$ of the differential equation
\begin{equation}
    \frac{d w}{dr} = \frac{4 \pi G}{c^2} \frac{(\mathcal{E} + P) (4+w) r}{c^2 - 2 G \mathcal{M}/r} - \frac{w}{r} (3+w) \ ,
    \label{eq:dwdr}
\end{equation}
with the boundary condition $w(r=0)=0$.
In the left panel of Fig.~\ref{fig:I14_LDHD} we display the posterior distribution for the normalised moment of inertia of a $1.4 M_\odot$ NS, $I_{1.4}/(MR^2)$ when the LD+HD filters are applied and the piecewise polytrope is matched at $n_{\rm match} = 0.16$~fm$^{-3}$ (blue shaded area).
Similarly to what is observed for the radius (see Fig.~\ref{fig:R14_016}), the use of a unique EoS (dash-dotted curves) slightly shifts the average value; in the case where the unique EoS is employed up to $0.16$~fm$^{-3}$ (blue dashed curve), the width of the distribution is considerably reduced.
The same trend is seen when the filter from GW170817 and NICER is included in the posterior (orange curves).
The impact of the unique EoS model on the uncertainties in the prediction is even larger, as expected, as far as crustal quantities are concerned.
This can be seen in the right panel of Fig.~\ref{fig:I14_LDHD}, showing the fractional moment of inertia of the crust of a $1.4 M_\odot$ NS.
In this case, the width of the distribution is about an order of magnitude smaller when only one EoS model is considered below the polytrope matching density.

The medians and 68$\%$ confidence intervals for NS radii, tidal deformabilities, and moment of inertia are summarised for the different treatments of the low-density EoS (crust) in Table~\ref{tab:medians}. The difference in medians with all filters applied between the unique low-density EoS and the full treatment is of the order of a few percent with larger values for the higher matching density. 
As discussed before, uncertainties are comparable for radii and tidal deformabilities, whereas the unique low-density EoS leads to a considerable reduction for the uncertainties on moments of inertia. 
Even though the difference in treatment is much smaller than the measurement uncertainty expected for current ground-based GW detectors and their upgrades, it is of the same order as the one expected for third-generation detectors \citep{Chatziioannou2021,Huxford2023,Iacovelli2023}. 
This means that the inference of NS radii and the NS EoS from the data has to handle the treatment of the inhomogeneous part of the EoS in the crust in a consistent way for the upcoming generation of detectors, whereas using a reasonable unique crust, such as the DH(SLY4) model, below its crust--core transition density, should be a good approximation for the current detectors.

As a note of caution, it should be mentioned here that the small difference between the unique crust and the full difference is largely influenced by the filter imposed on the EoS by the chiral-EFT results which strongly constrain the EoS for densities in the crust region. 
An underestimation of the systematic uncertainties inherent to the chiral-EFT calculations would mean that our results only give a lower limit on the effect of using a unique crust on the results. 

Concerning the NS moment of inertia, $I$, measurements are difficult, and for the moment no precise value exists. There is hope that the Square Kilometer Array (SKA) radio telescope project will be able to obtain it for the double-pulsar PSR J0737$-$3039A with a level of precision of $\lesssim 10$\% \citep{Watts2015}. 
Even in the fortunate event that another detected binary NS system provided a more favourable measurement of the moment of inertia, it seems probable, in view of the small differences in the treatment of the crust we obtained here, that using a unique crust model would not have any influence on the inference of NS properties from the moment of inertia, even with the SKA.

\begin{table*}[!ht]
    \centering
    \caption{Medians and $68\%$ confidence intervals of NS radii, dimensionless tidal deformabilities, moment of inertia, and crustal moment of inertia for a $M = 1.4\ M_\odot$ NS. 
    }
    \begin{tabular}{c | c c c c}
        \hline
            & $R_{1.4}$ [km] & $\Lambda_{1.4}$ & $I_{1.4}/(MR^2)$ & $I_{\rm crust, 1.4}/I_{1.4}$ \\
            \hline 
           Prior & $13.55^{+0.79}_{-1.63}$ & $902^{+413}_{-541}$ & $0.360^{+0.012}_{-0.010}$ & $0.018^{+0.039}_{-0.011}$ \\
           LD+HD & $13.57^{+0.54}_{-0.81}$ $\left(13.83^{+0.38}_{-0.75} \right)$ & $925^{+274}_{-332}$ $\left( 1025^{+226}_{-349} \right)$ & $0.360^{+0.005}_{-0.004}$ $\left( 0.358^{+0.001}_{-0.002} \right)$ & $0.045^{+0.012}_{-0.011}$ $\left( 0.045^{+0.004}_{-0.008} \right)$ \\
           LD+HD+LVC+NICER & $12.78^{+0.44}_{-0.40}$ $\left( 12.90^{+0.41}_{-0.42} \right)$ & $593^{+160}_{-118}$ $\left( 611^{+162}_{-133} \right)$ & $0.361^{+0.005}_{-0.005}$ $\left( 0.357^{+0.002}_{-0.002} \right)$ & $0.039^{+0.011}_{-0.009}$ $\left( 0.036^{+0.004}_{-0.004} \right)$ \\
           \hline
           Prior & $12.68^{+0.60}_{-0.91}$ & $544^{+195}_{-224}$ & $0.357^{+0.011}_{-0.010}$ & $0.016^{+0.032}_{-0.010}$ \\
           LD+HD & $12.70^{+0.42}_{-0.56}$ $\left(12.43^{+0.09}_{-0.19} \right)$ & $565^{+147}_{-154}$ $\left( 497^{+35}_{-65} \right)$ & $0.360^{+0.005}_{-0.003}$ $\left( 0.363^{+0.001}_{-0.002} \right)$ & $0.039^{+0.009}_{-0.009}$ $\left( 0.031^{+0.001}_{-0.001} \right)$ \\
           LD+HD+LVC+NICER & $12.62^{+0.36}_{-0.31}$ $\left( 12.46^{+0.07}_{-0.14} \right)$ & $540^{+120}_{-84}$ $\left( 505^{+27}_{-47} \right)$ & $0.361^{+0.005}_{-0.004}$ $\left( 0.364^{+0.001}_{-0.002} \right)$ & $0.039^{+0.008}_{-0.008}$ $\left( 0.031^{+0.001}_{-0.001} \right)$ \\
        \hline
    \end{tabular}
    \tablefoot{The piecewise polytrope is matched at $n_{\rm match} = 0.16$~fm$^{-3}$ (top lines) or at $n_{\rm match} = 0.32$~fm$^{-3}$ (bottom lines). Values in parentheses for the posteriors correspond to those obtained employing a unique EoS based on SLy4 until $n_{\rm match}$.}
    \label{tab:medians}
\end{table*}

\section{Conclusions}
\label{sec:conclusions}

In this paper, we propose a method to construct consistent and unified crust--core EoSs of cold and beta-equilibrated NSs for application to astrophysical analysis and inference. 
Starting from a beta-equilibrated relation between the energy density and the baryon number density, together with few nuclear empirical parameters, we show that it is possible to accurately reconstruct a unified and thermodynamically consistent EoS.
The associated numerical code \textsc{CUTER} has been tested and validated for use by the astrophysical community, and the first release is currently available.
This framework was validated on different nucleonic EoSs (see Sect.~\ref{sec:code-valid}), but is applicable to any beta-equilibrium (high-density) EoSs.

Within a Bayesian analysis and employing a piecewise polytrope at high density to represent the possible existence of non-nucleonic degrees of freedom and phase transitions in the NS core, we assessed the impact of the use of a unique instead of a unified and consistent EoS (crust) model.
Our results indicate that the use of a unique EoS model at lower densities underestimates the uncertainties in the prediction of global NS observables, thus highlighting the importance of employing a consistent EoS in inference schemes, which will be particularly important for next-generation detectors.

\begin{acknowledgements}
This work has been partially supported by the IN2P3 Master Project NewMAC, the ANR project `Gravitational waves from hot neutron stars and properties of ultra-dense matter' (GW-HNS, ANR-22-CE31-0001-01), the CNRS International Research Project (IRP) `Origine des \'el\'ements lourds dans l’univers: Astres Compacts et Nucl\'eosynth\`ese (ACNu)', the National Science Center, Poland grant 2018/29/B/ST9/02013 and Poland grant 2019/33/B/ST9/00942 and the National Science Foundation grant Number PHY 21-16686. This material is based upon work supported by NSF’s LIGO Laboratory which is a major facility fully funded by the National Science Foundation. The authors gratefully acknowledge the Italian Instituto Nazionale de Fisica Nucleare (INFN), the French Centre National de la Recherche Scientifique (CNRS), and the Netherlands Organization for Scientific Research for the construction and operation of the Virgo detector and the creation and support of the EGO consortium. The authors would like to thank N. Stergioulas and J. Novak for fruitful comments on the code.
\end{acknowledgements}

\bibliographystyle{aa}
\bibliography{biblio}
\end{document}